\documentclass[12pt]{article}
\usepackage{a4wide}
\usepackage{amsmath}
\usepackage{amssymb}
\usepackage{amsfonts}
\usepackage{array}
\usepackage{bm}
\usepackage{braket}
\usepackage{color}
\usepackage{comment}
\usepackage{graphicx}
\usepackage{mathrsfs}
\usepackage{wrapfig}

\newcommand{\clfn}{\setcounter{footnote}{0}}

\def\del{\partial}

\allowdisplaybreaks[1]

\begin{document}

\begin{titlepage}

    \begin{flushright}
      \normalsize TU-975\\
      \today
    \end{flushright}

\vskip1.5cm
\begin{center}
\large\bf\boldmath
A Scale-invariant Higgs Sector and
Structure of the Vacuum
\end{center}

\vspace*{0.8cm}
\begin{center}

{\sc K.~Endo and Y.~Sumino}\\[5mm]
  {\small\it Department of Physics, Tohoku University,}\\[0.1cm]
  {\small\it Sendai, 980-8578 Japan}

\end{center}

\vspace*{0.8cm}
\begin{abstract}
\noindent
In view of the current status of measured Higgs boson 
properties, we consider a question whether only the Higgs self-interactions can deviate significantly from the Standard-Model (SM) predictions.
This may be possible if the Higgs effective potential is irregular at the origin.
As an example we investigate an extended Higgs sector with singlet scalar(s) and classical scale invariance.
We develop a perturbative formulation necessary to analyze
this model in detail.
The behavior of
a phenomenologically valid potential in the perturbative regime is studied
around the electroweak scale.
We reproduce known results:
The Higgs self-interactions are substantially stronger than the SM predictions, while the Higgs interactions with other SM particles are barely changed.
We further predict that
the interactions of singlet scalar(s), which is a few to several times heavier than the Higgs boson, tend to be fairly strong.
If probed, these features will provide vivid clues to the structure of the vacuum.
We also examine Veltman's condition for the Higgs boson mass.

\vspace*{0.8cm}
\noindent

\end{abstract}

\vfil
\end{titlepage}

\section{Introduction}
It appears that
completion of the standard model  (SM) 
of particle physics, as it stands, has been achieved
with the discovery of the Higgs boson at the LHC experiments
\cite{Aad:2012tfa, Chatrchyan:2012ufa}.
In particular, the two parameters of the Higgs potential, 
\begin{align}
	V_\text{SM}(H)
=	-{\mu_\text{H}}^{2}H^{\dagger}H
	+\lambda_\text{H}
		\left(
			H^{\dagger}H
		\right)^{2} ,
\end{align}
namely, the vacuum expectation value (VEV) of the Higgs field $v_\text{H}=\mu_\text{H}/\sqrt{\lambda_\text{H}}=246~\text{GeV}$ and the Higgs boson mass $m_\text{h}=\sqrt{2}\mu_\text{H}\approx126~\text{GeV}$, are now determined.
Here, $H$ denotes the Higgs doublet field $H=(H^+,\,H^0)^\text{T}$.
This means that all the parameters of the SM have been determined.
After the discovery, investigations of properties of the Higgs boson are
being performed rapidly, such as measurements of its spin, $CP$, and 
couplings with various SM particles.
Up to now, there are no evident contradictions with the SM predictions.
Accuracies of the measurements are improving, and identification with the SM Higgs boson is becoming more likely.
These features, however, do not readily lead to the conclusion that the SM is confirmed altogether.
As an important piece, confirmation of the self-interactions of the Higgs boson is still missing, which is indispensable to unveil the structure of the Higgs potential.

There are some phenomenological and theoretical problems in the Higgs sector of the SM as of today.
For instance, there is a huge hierarchy between the electroweak and Planck scales, which generates the ``Naturalness problem''.
Furthermore, even if we permit fine tuning of parameters, we face a problem that the vacuum of the Higgs potential is unstable or metastable at a high-energy scale.\footnote{
The scale of the vacuum instability is very sensitive to the top quark mass and depends on our precise knowledge of the top quark mass in the future.
}
These problems give us motivations to consider extensions of the SM Higgs sector.

In view of the present status, it is an interesting question whether it is possible, with such an extension, that only the Higgs self-interactions are significantly different from the SM predictions, while other properties of the Higgs boson are barely affected.
Naively, one may expect that deviations of the Higgs self-couplings from the SM values can be expressed as effects of higher-dimensional operators, which are suppressed by a cut-off scale $\varLambda$ as $(v_\text{H}/\varLambda)^n$.
It is based on a general model-independent argument in the case that the Higgs effective potential can be expanded as a polynomial in the Higgs field $H$.
Since the cut-off scale may be of order a few
TeV scale or higher, one may expect that the deviation is already 
constrained to be quite small.

One direction to evade such an argument is to consider models with
non-decoupling effects to Higgs self-interactions.
There are many examples \cite{Kanemura:2002vm,Kanemura:2004mg}.
Here, we would like to consider a different possibility.
In various physics of spontaneous symmetry breakdown (including condensed matter physics), there appear effective potentials which cannot be expanded in polynomials in the field variables, namely effective potentials which are irregular at the origin.
For example, the theory of superconductivity at $T=0$ gives such an effective potential, and in certain strongly interacting systems such singular potentials are also expected to appear (although they are difficult to compute reliably).
As an example which is perturbatively computable within relativistic quantum field theory, we consider an extended Higgs sector with classical scale invariance.
In this case, typically the Higgs effective potential takes a form 
$\sim \lambda \phi^4\left[\ln\left(\phi^2/{v_\text{H}}^2\right)-1/2\right]$, which is irregular at the origin via the Coleman-Weinberg (CW) mechanism \cite{Coleman:1973jx}, where $\phi=\sqrt{2}\,\text{Re}\,H^0$.
If this potential is expanded about the VEV in terms of the physical Higgs field $h=\phi-v_\text{H}$, the expansion generates powers of $h$ higher than the quartic power, and they arise in powers of $h/v_\text{H}$ without being suppressed by a cut-off scale.
At the same time, it is expected that the triple and quartic couplings of $h$ can have order-unity deviations from the SM values.

In recent years, CW-type potentials have been studied extensively as models of an extended Higgs sector, with different motivations.
For instance, classical scale invariance as a solution to the hierarchy problem has become a hot subject.
These are models which become scaleless at a new physics scale $M$
(e.g.\ Planck scale), ${\mu_\text{H}}^{2}(M)=0$.
In these models, typically the scale invariance is broken radiatively by the CW mechanism, which (in steps) leads to a generation 
of the electroweak scale as the Higgs VEV \cite{Foot:2007as, Espinosa:2007qk, AlexanderNunneley:2010nw, Dermisek:2013pta, Masina:2013wja, Antipin:2013exa, Guo:2014bha, Tamarit:2014dua}, or generates scales involving nonzero VEVs of other scalar fields \cite{Hempfling:1996ht, Chang:2007ki, Meissner:2006zh, Foot:2007iy, Iso:2009ss, Iso:2012jn, Englert:2013gz, Hambye:2013dgv, Carone:2013wla, Gabrielli:2013hma, Abel:2013mya, Radovcic:2014rea, Khoze:2014xha, Altmannshofer:2014vra, Benic:2014aga}.
Even without explicit realizations, classical scale invariance is often
mentioned as a possibility or guideline for an underlying mechanism.


The purpose of this paper is to explore physics in the vicinity of our vacuum, characterized by the effective potential of an extended Higgs sector.
As a minimal extension of the SM Higgs sector, we consider a scale-invariant Higgs potential with a real singlet scalar field which belongs to the $N$ representation of a global $O(N)$ symmetry.
Though there have been several works in which the same or similar 
extensions are considered, we re-analyze the model by examining the
validity range of perturbative calculations of the effective potential 
in detail.
We can obtain a phenomenologically valid potential within the validity range of a perturbative analysis.
We clarify its behavior around the electroweak scale.
We also examine Veltman's condition \cite{Veltman:1980mj}
for the Higgs mass, 
which is a criterion for judging a ``Naturality''
of the scalar potential by examining the coefficient of the quadratic
divergence.

Some of previous works on similar models are subject to possible instability
of their predictions when higher-order perturbative
corrections are included.
Such instability was already pointed out by the original CW paper,
and we re-examine this condition.
Other works, which are legitimate with respect to the perturbative
validity, use the framework of Gildener-Weinberg (GW) \cite{Gildener:1976ih}
to analyze the effective potential.
In this framework, one concentrates on a one-dimensional
subspace of the configuration space of the effective potential,
namely that corresponding to the physical Higgs direction, and examines the
potential shape on that subspace.
On the other hand, we analyze the effective potential in a way
closer to the original CW approach, which enables to
clarify a global structure of
the potential shape in the configuration space.

Ref.~\cite{Dermisek:2013pta} analyzed a gauged and non-gauged version
of a scale-invariant model, and the latter is close to the one
we examine.
Phenomenologically we reproduce similar aspects, hence
we state the
differences of our analysis in comparison.
As mentioned above, Ref.~\cite{Dermisek:2013pta} uses the GW 
framework to analyze the properties of the effective potential.
Instead we present a formulation which enables analysis
of global properties of the effective potential.
As a result, for instance,
we are able to analyze the structure of the potential
in every direction around the vacuum in the configuration space and
predict the interactions among the physical scalar particles.

This paper is organized as follows.
In Sec.~2 we explain our model.
In Sec.~3, we develop a theoretical framework needed for a perturbative analysis of the model.
We give results of our numerical analysis in Sec.~4.
Conclusions and discussion are given in Sec.~5.

\section{Model and effective Higgs potential}
\subsection{Lagrangian}

The scale-invariant limit of the SM has long been excluded experimentally.
Hence, to impose classical scale invariance, we need to extend the Higgs sector.
As a minimal extension, we consider a scale-invariant extension of the SM with an additional real singlet scalar field with a Higgs-portal coupling.
We consider the case where the singlet scalar field is in the fundamental representation of a global $O(N)$ group:
$	\vec{S}
=	(S_{1},	\cdots,S_{N})^\text{T}
$.
The Lagrangian is given by
\begin{align}
	\mathscr{L}
=&\,
	\left[
		\mathscr{L}_\text{SM}
	\right]_{\mu_\text{H}\rightarrow 0}
	+\frac{1}{2}(\del_{\mu}{\vec{S}})^{2}
	-{\lambda_\text{HS}}
		(H^{\dagger}H)(\vec{S}\cdot\vec{S})
	-\frac{\lambda_{S}}{4}(\vec{S}\cdot\vec{S})^2\,,
\label{eq:our_model}
\end{align}
where the real singlet field $\vec{S}$ interacts with itself and the Higgs doublet field $H$ via the self-interaction and portal interaction with the coupling constants $\lambda_\text{S}$ and $\lambda_\text{HS}$, respectively.

\subsection{Effective potential up to one-loop level}

The one-loop effective potential in the Landau gauge, renormalized in the $\overline{\text{MS}}$-scheme, is 
given by 
\begin{align}
&	V_\text{eff}(\phi,\,\varphi)
=	V_\text{tree}(\phi,\,\varphi)
	+V_\text{1-loop}(\phi,\,\varphi)\,,
\label{eq:effective_pot_unimp_complete}
\\
&	V_\text{tree}(\phi,\,\varphi)
=	\frac{\lambda_\text{H}}{4}{\phi}^4
	+\frac{\lambda_\text{HS}}{2}{\phi}^2{\varphi}^2
	+\frac{\lambda_\text{S}}{4}{\varphi}^2\,,
\label{eq:effective_pot_unimp_tree-level}
\\
&	V_\text{1-loop}(\phi,\,\varphi)
=	\sum_{i}
	\frac{n_i}{4(4\pi)^2}{M_i}^4(\phi,\,\varphi)
	\left[
		\ln
			\frac{{M_i}^2(\phi,\,\varphi)}{\mu^2}
		-c_i
	\right]\,.
\label{eq:effpot_ms-bar_mod}
\end{align}
Here, the expectation values of the scalar fields in the presence of source $J$ are given by
\begin{align}
	\braket{H}_J
=	\frac{1}{\sqrt{2}}
	\left(
	\begin{array}{c}
		0	\\
		\phi
	\end{array}
	\right)\,,
~~~~~
	\braket{\vec{S}}_J
=		( \varphi	,0	,\cdots	,0	)^\text{T}
~~;~~~
\phi,~\varphi \in \mathbb{R}
 \,.
\end{align}
The index $i$ denotes the internal particle in the loop, and their parameters are given by\footnote{
Strictly speaking, the terminology ``Nambu-Goldstone (NG) bosons'' for the internal particles is inadequate except at the vacuum configuration and depends on how the symmetries are broken by the vacuum.
More precisely, ``NG bosons'' represent the scalar modes which are orthogonal to the radial directions of $H$ and $S$.}
\begin{align}
&\text{$W$ bosons: }
	n_W=6\,,~~
	{M_W}^2
	=\frac{1}{4}g^2{\phi}^{2}\,,~~
	c_W=\frac{5}{6}\,;
\nonumber\\
&\text{$Z$ boson: }
	n_Z=3\,,~~
	{M_Z}^2
		=\frac{1}{4}(g^2+{g'}^2){\phi}^{2}\,,~~
	c_Z=\frac{5}{6}\,;
\nonumber\\
&\text{massive scalar bosons: }
	n_\pm=1\,,~~
	{M_\pm}^2
		=F_\pm\,,~~
	c_\pm=\frac{3}{2}\,;
\nonumber\\
&\text{NG bosons of the Higgs field: }
	n_\text{NG}=3\,,~~
	{M_\text{NG}}^2
		=	\lambda_\text{H}{\phi}^2
			+\lambda_\text{HS}{\varphi}^2\,,~~
	c_\text{NG}=\frac{3}{2}\,;
\nonumber\\
&\text{NG bosons of the singlet field: }
	n_\text{NG}=N-1\,,~~
	{M_\text{NG}}^2
		=	\lambda_\text{HS}{\phi}^2
			+\lambda_\text{S}{\varphi}^2\,,~~
	c_\text{NG}=\frac{3}{2}\,;
\nonumber\\
&\text{up-type quarks ($F=u,\,c,\,t$): }
	n_F=-4N_\text{C}\,,~~
	{M_F}^2
		=\frac{1}{2}\left(y^{(\text{U})}_{FF}\right)^{2}
			{\phi}^{2}\,,~~
	c_F=\frac{3}{2}\,;
\nonumber\\
&\text{down-type quarks ($f=d,\,s,\,b$): }
	n_f=-4N_\text{C}\,,~~
	{M_f}^2
		=\frac{1}{2}\left(y^{(\text{D})}_{ff}\right)^{2}
			{\phi}^{2}\,,~~
	c_f=\frac{3}{2}\,;
\nonumber\\
&\text{charged leptons ($f=e,\,\mu,\,\tau$): }
	n_f=-4\,,~~
	{M_f}^2
		=\frac{1}{2}\left(y^{(\text{E})}_{ff}\right)^{2}
			{\phi}^{2}\,,~~
	c_f=\frac{3}{2} ,
	\label{internal-modes}
\end{align}
and $F_\pm$ are defined by
\begin{align}
	F_{\pm}(\phi,\,\varphi)
=&	\frac{3\lambda_\text{H}+\lambda_\text{HS}}{2}{\phi}^2
	+\frac{\lambda_\text{HS}+3\lambda_\text{S}}{2}{\varphi}^2
\nonumber\\
&	\pm
	\sqrt{
		\left(
			\frac{3\lambda_\text{H}-\lambda_\text{HS}}{2}
				{\phi}^2
			+\frac{\lambda_\text{HS}-3\lambda_\text{S}}{2}
				{\varphi}^2
		\right)^2
		+4{\lambda_\text{HS}}^2{\phi}^2{\varphi}^2
	}\,.
\label{eq:F-factor_def}
\end{align}
Hereafter, we neglect all the Yukawa couplings except the top Yukawa coupling, $y_\text{t}=y^{(\text{U})}_{33}\sim \mathcal{O}(1)$, since other Yukawa couplings are fairly small.
It is customary to denote
the summation over $i$ together with particle's statistical factor 
as supertrace ``$\text{STr}$,''
which we also use below.

\subsection{Renormalization group analysis}
\label{sec:rg_analysis}
We can extend the applicability range of the effective potential by a renormalization-group (RG) improvement.
According to the general formulation \cite{Bando:1992np,Bando:1992wy}, the ($L+1$)-loop beta functions and anomalous dimensions can be used to improve the $L$-loop effective potential.
Here, we obtain the leading-logarithmic (LL) potential by improving the tree-level potential by the
one-loop beta functions and anomalous dimensions.
It is argued in \cite{Bando:1992np,Bando:1992wy} that, within the range where $\ln({M_{i}}^{2}/\mu^2)$ are not too large, combining the one-loop effective potential with the
one-loop beta functions and anomalous dimensions gives a better approximation.
Hence, we obtain an improved-next-to-leading-order (improved-NLO) potential by combining them.
Comparing NLO, LL and improved-NLO potentials, we can examine validity
(stability) of the predictions in the vicinity of the vacuum.

The beta functions and anomalous dimensions are defined by
\begin{align}
&	\beta_{X}
=	\mu\frac{d X}{d \mu}\,,
~~~~~
	\gamma_{A}
=	\frac{\mu}{A}\frac{d A}{d \mu}\,,
\end{align}
respectively, where $X$ is a coupling constant and $A$ is a field, both renormalized at scale $\mu$.
The one-loop beta functions and anomalous dimensions of the model are given by
\begin{align}
&	\beta_{\lambda_\text{H}}
=	\frac{1}{(4\pi)^2}
	\left[
		\frac{9}{8}g^4
		+\frac{3}{4}g^2 {g'}^2
		+\frac{3}{8}{g'}^4
		-6{y_\text{t}}^4
		\right.
\nonumber\\
&\hspace{30mm}		
		\left.
		-4\left(
			\frac{9}{4}g^2
			+\frac{3}{4}{g'}^2
			-3{y_\text{t}}^2
		\right)\lambda_\text{H}
		+24{\lambda_\text{H}}^2
		+2N{\lambda_\text{HS}}^2
	\right]\,,
\\
&	\beta_{\lambda_\text{HS}}
=	\frac{1}{(4\pi)^2}
	\lambda_\text{HS}
	\left[
		-2\left(
			\frac{9}{4}g^2
			+\frac{3}{4}{g'}^2
			-3{y_\text{t}}^2
		\right)
		+12{\lambda_\text{H}}
		+8\lambda_\text{HS}
		+2(N+2)\lambda_\text{S}
	\right]\,,
\label{eq:MN_beta-func_portal}
\\
&	\beta_{\lambda_\text{S}}
=	\frac{1}{(4\pi)^2}
	\left(
		8{\lambda_\text{HS}}^2
		+2(N+8){\lambda_\text{S}}^2
	\right)\,,
\\
&	\gamma_{\phi}
=	\frac{1}{(4\pi)^2}
	\left(
		\frac{9}{4}g^2
		+\frac{3}{4}{g'}^2
		-3{y_\text{t}}^2
	\right)\,,
\\
&	\gamma_{\varphi}
=	0\, .
\label{eq:MN_anomalous_dim_singlet}
\end{align}
The beta functions which do not include $\lambda_\text{HS}$ or $\lambda_\text{S}$ are suppressed.

We obtain the improved potentials by taking the renormalization scale as $t=\ln(\sqrt{{\phi}^2+{\varphi}^2}/v_\text{H})$.
\begin{align}
&	V_\text{eff}^{(\text{LL})}(\phi,\,\varphi)
=	\frac{\lambda_\text{H}(t)}{4}{\phi}^4(t)
	+\frac{\lambda_\text{HS}(t)}{2}
		{\phi}^2(t){\varphi}^2(t)
	+\frac{\lambda_\text{S}(t)}{4}{\varphi}^4(t)\,,
\label{eq:VLL}
\\
&	V_\text{eff}^{(\text{\scriptsize imp-NLO})}(\phi,\,\varphi)
=	\frac{\lambda_\text{H}(t)}{4}{\phi}^4(t)
	+\frac{\lambda_\text{HS}(t)}{2}
		{\phi}^2(t){\varphi}^2(t)
	+\frac{\lambda_\text{S}(t)}{4}{\varphi}^4(t)
\nonumber\\
&~~~~~~~~~~~~~~~~~~~~~~~~~
	+\sum_{i}
	\frac{n_i}{4(4\pi)^2}{M_i}^4\left(\phi(t),\,\varphi(t)\right)
	\left[
		\ln
			\frac{{M_i}^2\left(\phi(t),\,\varphi(t)\right)}{\mu^2(t)}
		-c_i
	\right]\,,
\label{eq:effpot_ms-bar_mod_improved}
\end{align}
where
\begin{align}
&	\mu(t)
=	v_\text{H}\,e^{t}\,,
\\
&	\phi(t)
=	\xi_\phi(t)\phi\,,
\\
&	\varphi(t)
=	\xi_\varphi(t)\varphi\,,
\\
&	\xi_i(t)
=	\exp
		\left[
			\int^{t}_{0}\gamma_i(t')dt'
		\right] .
\end{align}

\section{Perturbatively valid parameter region}

In a previous work \cite{Meissner:2006zh} a parameter region has been searched assuming
\begin{align}
	\lambda_\text{H}
\sim
	\lambda_\text{HS}
<	1\,.
\end{align}
In that case, however, the Coleman-Weinberg mechanism does not work properly, since the one-loop corrections cannot compete against the tree-level terms in such a parameter region if the order counting in perturbation theory is legitimate.
(This is in analogy to the case of pure scalar $\phi^4$ field theory considered in the original CW paper \cite{Coleman:1973jx}.)
As a result, the renormalization scale considered in \cite{Meissner:2006zh} is about $10^{5}v_\text{H}$, where the large logarithmic corrections invalidate a perturbative analysis.
In this section, we reconsider the parameter region, in which the results are perturbatively valid and the CW mechanism works properly.

As mentioned in the Introduction, 
there is a framework of analysis to find a vacuum of a one-loop effective potential and to compute particle contents of scalar bosons 
systematically, which was introduced by Gildener and Weinberg 
\cite{Gildener:1976ih}.
In that framework we obtain the following general results.
A perturbatively valid vacuum appears on the ray of a flat direction of the tree-level potential;
a light scalar boson ``scalon'' (corresponding to
the Higgs boson in our context)
appears in addition to heavy scalar bosons;
interactions among the scalon and the other scalar bosons are derived.
Although it is a useful framework, we consider that
our framework presented below is advantageous to analyze 
more global features of the potential, including the vicinity of the vacuum.

\subsection{Order counting in perturbative expansion}
\label{s3.1}

We investigate an order counting 
among the coupling constants, with which the CW mechanism is 
expected to work in the perturbative regime.
To realize the CW mechanism,
the tree-level and one-loop contributions to the
effective potential should be comparable and compete
with each other.
Parametrically this requires a relation
\begin{align}
	|\lambda_\text{H}|
\sim
	\frac{N\,{\lambda_\text{HS}}^2}{(4\pi)^2}
	-\frac{N_\text{C}\,{y_\text{t}}^4}{(4\pi)^2}
\ll 1
\label{eq:MN_order-count_01}
\end{align}
to be satisfied.
(This is in analogy to the case of massless scalar QED considered in the original CW paper \cite{Coleman:1973jx}, in which $\lambda\sim\frac{e^4}{(4\pi)^2}\ll 1$ is required as a consistent parameter region.)
Note that we take $y_\text{t}$ into account in Eq.~(\ref{eq:MN_order-count_01}) since the top quark gives the dominant one-loop contribution among the SM particles which couple to the Higgs particle.
According to Eq.~(\ref{eq:MN_order-count_01}), we consider that $\lambda_\text{H}$ and ${\lambda_\text{HS}}^2$ 
(as well as ${y_\text{t}}^4$) are naively counted as the same order quantities in perturbative expansions.\footnote{
We do not consider a fine cancellation between the ${\lambda_\text{HS}}^2$
and ${y_\text{t}}^4$ contributions.
}
It follows that the relation between $\lambda_\text{H}$ and ${\lambda_\text{HS}}$ should read 
\begin{align}
	|{\lambda_\text{H}}|
\ll	|{\lambda_\text{HS}}|\,.
\label{eq:MN_order-count_02}
\end{align}
Thus, Eq.~(\ref{eq:MN_order-count_01}) and Eq.~(\ref{eq:MN_order-count_02}) indicate that ${\lambda_\text{H}}^2$ and ${\lambda_\text{HS}}^2$ need to be counted as different orders, although they both belong to the one-loop contributions.
Furthermore, ${\lambda_\text{HS}}$ needs to be large, at least of order $\sqrt{N_\text{C}/N}\,y_\text{t}
\approx 1.7\,N^{-1/2}$, in order to beat the top quark negative contribution, for stabilizing the vacuum.

We derive a systematic approximation
of the effective potential Eq.~(\ref{eq:effpot_ms-bar_mod}), taking into account the above order counting. 
First, the contributions from the NG bosons of the Higgs field on the right-hand-side of Eq.~(\ref{eq:effpot_ms-bar_mod}) can be written as follows:
\begin{align}
&	\frac{3}{64\pi^{2}}
	(
		\lambda_\text{H}{\phi}^2
		+\lambda_\text{HS}{\varphi}^2
	)^2
	\left[
		\ln
		\left(
			\frac{
				\lambda_\text{H}{\phi}^2
				+\lambda_\text{HS}{\varphi}^2
			}{\mu^2}
		\right)
		-\frac{3}{2}
	\right]
\nonumber\\
&~~~~~~~~~~~~~~
\simeq
	\frac{3}{64\pi^{2}}
	\left(
		{\lambda_\text{HS}}{\varphi}^2
	\right)^2
	\left[
		\ln
		\left(
			\frac{\lambda_\text{HS}{\varphi}^2}{\mu^2}
		\right)
		-\frac{3}{2}
	\right]\,.
\label{eq:3rd_term}
\end{align}
Secondly, $F_{\pm}(\phi,\,\varphi)$ in the fourth and fifth terms in Eq.~(\ref{eq:effpot_ms-bar_mod}) can be written as follows:
\begin{align}
		F_{\pm}(\phi,\,\varphi)
\simeq&	
	\frac{\lambda_\text{HS}}{2}{\phi}^2
	+\frac{\lambda_\text{HS}+3\lambda_\text{S}}{2}{\varphi}^2
\nonumber\\
&	\pm
	\sqrt{
		\left[
			-\frac{\lambda_\text{HS}}{2}{\phi}^2
			+\frac{\lambda_\text{HS}-3\lambda_\text{S}}{2}{\varphi}^2
		\right]^2
		+4{\lambda_\text{HS}}^2{\phi}^2{\varphi}^2
	}
\nonumber\\
\equiv&
	F_{\pm\text{app}}(\phi,\,\varphi)\,,
	\label{Fapp}
\end{align}
where we have defined $F_{\pm\text{app}}(\phi,\,\varphi)$ as the approximate form of $F_{\pm}(\phi,\,\varphi)$.
Then we substitute Eqs.~(\ref{eq:3rd_term}) and (\ref{Fapp}) to the expression of the effective potential Eq.~(\ref{eq:effpot_ms-bar_mod}).

\begin{figure}[tbp]
\begin{center}
\includegraphics[width=7cm]{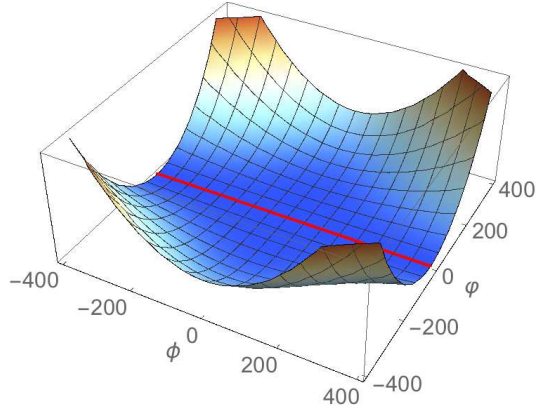}
\end{center}
\caption{\small The LO effective potential $V_\text{LO}$ as a function of $\phi$ and $\varphi$. 
The red line shows the $\phi$ axis, which composes the minima of the potential.
\label{fig:VLO}}
\end{figure}

We also have to apply Eq.~(\ref{eq:MN_order-count_01}) or Eq.~(\ref{eq:MN_order-count_02}) to the tree-level potential in Eq.~(\ref{eq:effective_pot_unimp_tree-level}).
In particular, Eq.~(\ref{eq:MN_order-count_01}) can be interpreted as follows.
Although the term proportional to $\lambda_\text{H}$ is tree-level, after taking into account the above order counting of the coupling constants, this term should be regarded as next-to-leading order (NLO), in contrast to the term proportional to $\lambda_\text{HS}$, which is at the LO.
For this reason, the LO contributions to the effective potential in the above order counting is given by
\begin{align}
	V_\text{LO}
=	\frac{\lambda_\text{HS}}{2}{\phi}^2{\varphi}^2
	+\frac{\lambda_\text{S}}{4}{\varphi}^4 .
\label{eq:0th_perturbation_order}
\end{align}
We show $V_\text{LO}$ in Fig.~\ref{fig:VLO}.
At LO, the potential is flat along the $\phi$ axis, which composes the minima of this potential.\footnote{
If $\lambda_\text{S}$ is small and
of the same order as $\lambda_\text{H}$, $\lambda_\text{S}{\varphi}^4$ 
should be counted as NLO.
In this case, both $\phi$ and $\varphi$ axes become the flat
minima of $V_\text{LO}$.
Other features, especially the results presented in the next section,
are hardly affected, if the global minimum of
$V_\text{LO}+V_\text{NLO}$ is on the
$\phi$ axis.
}
Thus, at LO, the vacuum is not determined uniquely.
At every vacuum the Higgs boson is massless, while the singlet scalers are massive.  

The NLO contributions are given by
\begin{align}
	V_\text{NLO}
=&	\frac{\lambda_\text{H}}{4}{\phi}^4
	+\frac{
			{F_{+\text{app}}}^{2}(\phi,\,\varphi)
		}{64\pi^2}
		\left[
			\ln
				\left(
					\frac{F_{+\text{app}}
							(\phi,\,\varphi)
						}{\mu^{2}}
				\right)
			-\frac{3}{2}
		\right]
\nonumber\\
&	+\frac{3}{64\pi^{2}}
		\left(
			{\lambda_\text{HS}}{\varphi}^2
		\right)^2
		\left[
			\ln
				\left(
					\frac{\lambda_\text{HS}{\varphi}^2}{\mu^2}
				\right)
			-\frac{3}{2}
	\right]
\nonumber\\
&+\frac{N-1}{64\pi^{2}}
	\left(
		\lambda_\text{HS}{\phi}^2
		+\lambda_\text{S}{\varphi}^2
	\right)^2
	\left[
		\ln
		\left(
			\frac{
				\lambda_\text{HS}{\phi}^2
				+\lambda_\text{S}{\varphi}^2
			}{\mu^2}
		\right)
		-\frac{3}{2}
	\right]
\nonumber\\
&	-\frac{4N_\text{C}}{64\pi^2}
	{M_\text{t}}^{4}(\phi)
	\left[
		\ln
			\left(
				\frac{{M_\text{t}}^{2}(\phi)}{\mu^2}
			\right)
		-\frac{3}{2}
	\right]
\nonumber\\
&	+\frac{6}{64\pi^{2}}
	{M_W}^4(\phi)
	\left[
		\ln
		\left(
			\frac{{M_W}^2(\phi)}{\mu^2}
		\right)
		-\frac{5}{6}
	\right]
\nonumber\\
&	+\frac{3}{64\pi^{2}}
	{M_Z}^4(\phi)
	\left[
		\ln
		\left(
			\frac{{M_Z}^2(\phi)}{\mu^2}
		\right)
		-\frac{5}{6}
	\right] .
\label{eq:1st_perturbation_order}
\end{align}
We explain the reason for omitting $F_{-\text{app}}$ 
in the next subsection.

As seen above, at every vacuum of $V_\text{LO}$, $\varphi=0$ and 
the $\varphi$-direction becomes a massive mode.
Hence, from consistency of the perturbative expansion, we expect that $\varphi=0$ holds also at the vacuum of $V_\text{LO}+V_\text{NLO}$.
According to the general argument on the potential with a flat direction at LO, 
with an appropriate choice of the couplings $(\lambda_\text{H},\,\lambda_\text{HS},\,\lambda_\text{S})$, $V_\text{LO}+V_\text{NLO}$ exhibits a minimum on the $\phi$-axis by the CW mechanism, and the Higgs boson becomes massive at NLO.
Hence, the Higgs boson is generally expected to be much lighter than the singlet scalars.
Thus, the SM gauge group is broken by the vacuum as required $SU(2)_\text{L}\times U(1)_\text{Y}\to U(1)_\text{EM}$, while the $O(N)$ global symmetry is unbroken. [There appears no NG mode with respect to the $O(N)$ group.]

\subsection{Comment on the Hessian matrix}
\clfn

A special prescription is needed for computing the CW potential, in the case that the minimum of the effective  potential is not determined uniquely at the LO of the perturbative expansion, such as in Eq.~(\ref{eq:0th_perturbation_order}).

Generally the arguments of logarithms in a one-loop effective potential include the eigenvalues ${m_i}^2$ of the Hessian matrix of the tree-level scalar potential.
The Hessian matrix is a matrix whose elements are given by $\frac{\del^2 V_\text{tree}}{\del x_i \del x_j}$, where $x_i$ denotes a scalar field.
Therefore, the eigenvalues ${m_i}^2$ at the potential minimum coincide with the mass-squared eigenvalues of the scalar fields at tree level.
In the case that the tree-level potential has a unique minimum, all the eigenvalues of the Hessian matrix at the minimum are positive, ${m_i}^2> 0$, and ${m_i}^4\ln {m_i}^2$ in the one-loop effective potential are well-defined.\footnote{
For massless modes, such as NG bosons of the Higgs field, we define the values of ${m_i}^4\ln {m_i}^2$ in the limit ${m_i}^2\rightarrow 0$, i.e., zero.
See Eq.~(\ref{eq:3rd_term}).
}

Since in our case $V_\text{LO}$ is flat along the $\phi$-axis, ${m_\phi}^2$ (the eigenvalue in the direction of $\phi$) can be negative at configuration points $(\phi,\,\varphi)$ infinitesimally away from the $\phi$-axis.
Instead, if we determine the potential minimum of $V_\text{LO}+V_\text{NLO}$ and compute the Hessian matrix of $V_\text{LO}+V_\text{NLO}$ in the vicinity of the minimum, all the eigenvalues are positive.
Thus, a naive perturbative treatment is inappropriate in computing quantum fluctuations in the vicinity of a vacuum in the case that there is a negative eigenvalue.

Here we adopt a prescription\footnote{
This corresponds to the following prescription for computing $V_\text{eff}$ by the background field method.
We determine the propagator of the relevant quantum field not by the LO vertices alone but also by including one-loop self-energy corrections.
Note that both LO vertices and one-loop self-energy corrections are dependent on the background fields $(\phi,\,\varphi)$ and determined from $V_\text{LO}$ and $V_\text{NLO}$, respectively.
} that (only) the eigenvalue ${m_\phi}^2$ cannot be determined by $V_\text{LO}$, and that this eigenvalue should be determined by $V_\text{LO}+V_\text{NLO}$, whose effect through ${m_\phi}^4\ln {m_\phi}^2$ should be included in $V_\text{NNLO}$.\footnote{
This is consistent in the vicinity of the $\phi$-axis since ${m_\phi}^2$ is an NLO quantity.
}
Since we do not compute $V_\text{NNLO}$, in practice we simply neglect its contribution.
The other eigenvalue ${m_\varphi}^2$ determined by $V_\text{LO}$ is positive.
In Eq.~(\ref{internal-modes}), the two eigenvalues are given by $F_\pm(\phi,\,\varphi)$.
The eigenvalues of the modes orthogonal to these radial modes in the scalar sector are composed by those of the three degenerate modes of $H$ and those of the $N-1$ degenerate modes of $\vec{S}$, all of which are non-negative; see Eqs.~(\ref{internal-modes}) and (\ref{eq:3rd_term}).\footnote{
At the vacuum, the three modes orthogonal to the radial mode of $H$ are identified with the NG modes, whose eigenvalues vanish to all orders, while all the $N$ modes of $\vec{S}$ become degenerate, since the $O(N)$ symmetry is unbroken.
}
We consider that these should be included in the computation of $V_\text{NLO}$.
Based on these considerations, we omit the contribution of $F_{-\text{app}}$ (the eigenvalue in the direction $\phi$) in Eq.~(\ref{eq:1st_perturbation_order}).

An analysis for a consistent treatment of
the NG modes has been performed recently \cite{Martin:2014bca},
which is similar in spirit to the above argument.
We will further develop the above method for including higher-order
corrections consistently in our future work.

\subsection{Relation between $\lambda_\text{H}$ and $\lambda_\text{HS}$ at $\mu=v_\text{H}$}
\label{S3.3}
\clfn

To study the effective potential in the vicinity of the vacuum, a natural choice of the renormalization scale would be $\mu=v_\text{H}$, where $v_\text{H}= 246$~GeV is the VEV of the Higgs field.
Here, we set $\mu=v_\text{H}$ and $\varphi=0$ and examine a relation among the scalar couplings $(\lambda_\text{H},\,\lambda_\text{HS},\,\lambda_\text{S})$. (Other couplings are fixed to the SM values.)
As a result we obtain a relation between $\lambda_\text{H}$ and $\lambda_\text{HS}$.

Setting $\varphi=0$, the effective potential takes a form
\begin{align}
	V_\text{eff}(\phi,\,\varphi=0)
=	C_1{\phi}^4
	+C_2{\phi}^4
		\ln
		\left(
			\frac{\phi}{\mu}
		\right),
\end{align}
where $C_1$ and $C_2$ are constants dependent on $\lambda_\text{H}$, $\lambda_\text{HS}$ and independent of $\lambda_\text{S}$.
This shows that both mass and VEV of the Higgs boson are independent of $\lambda_\text{S}$.
[This is not the case if we use RG-improved potentials, since terms including both $\lambda_\text{S}$ and $\ln(\sqrt{{\phi}^2+{\varphi}^2}/v_\text{H})$ are resummed.]

\begin{figure}[tbp]
\begin{center}
\includegraphics[width=9cm]{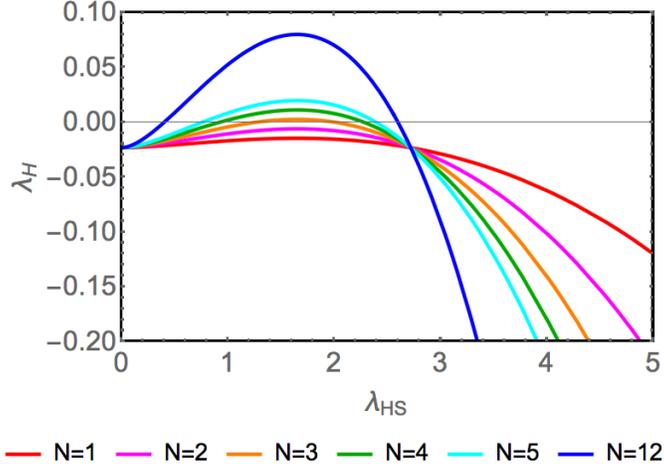}
\end{center}
\caption{\small 
The relation between $\lambda_\text{H}$ and $\lambda_\text{HS}$ obtained at the vacuum ($\varphi=0$) with a choice of the renormalization scale $\mu=v_\text{H}$, for various $N$. [Eq.~(\ref{eq:relation_self_and_potal})].
\label{fig:why_cannot_mu=246gev}
}
\end{figure}

The potential minimum is determined by
\begin{align}
	\frac{\del}{\del \phi}
		V_\text{eff}(\phi,\,\varphi=0)	
		\biggr|_{\phi=v_\text{H}}
=	0 ,
\end{align}
from which we obtain $v_\text{H}$ as a function of $\lambda_\text{H}$, $\lambda_\text{HS}$ and $\mu$.
Then we set $\mu=v_\text{H}$:
\begin{align}
	v_\text{H}
=	f(\lambda_\text{H},\,\lambda_\text{HS};\,\mu)
\biggr|_{\mu=v_\text{H}}
\,.
\label{eq:relation_self_and_potal}
\end{align}
Since $v_\text{H}$ is the only dimensionful parameter, the right-hand-side is proportional to $v_\text{H}$ and we can divide both sides by $v_\text{H}$.
This gives a relation between $\lambda_\text{H}$ and $\lambda_\text{HS}$, which are renormalized at $\mu=v_\text{H}$.
The relation is shown in Fig.~\ref{fig:why_cannot_mu=246gev}, obtained by solving Eq.~(\ref{eq:relation_self_and_potal}) numerically.\footnote{
If we neglect the gauge couplings (and all the Yukawa couplings except $y_\text{t}$),
we obtain a simple relation:
\begin{align}
	\lambda_\text{H}
=	-\frac{1}{16\pi^2}
		[
			\,{y_\text{t}}^4\,(3+3\ln 2-6\ln y_\text{t})
			+N\,{\lambda_\text{HS}}^2(\ln \lambda_\text{HS}-1)
		].
\end{align}
This gives a good approximation of the numerical result.
}
Note that $\lambda_\text{HS}$ should be of order $\sqrt{N_\text{C}/N}\,y_\text{t}\approx 1.7\,N^{-1/2}$ or larger (see Sec.~\ref{s3.1}), while it should be smaller than order $4\pi$ to ensure perturbativity.

\section{Results of analysis}

\subsection{Phenomenologically valid parameters}

Using the effective potential $V_\text{LO}+V_\text{NLO}$ obtained in the previous section, we search for a phenomenologically valid parameter region for the couplings $(\lambda_\text{H},\,\lambda_\text{HS},\,\lambda_\text{S})$.
We require that the observed mass and VEV of the Higgs boson are reproduced.

The analysis in Sec.~\ref{S3.3} shows that, as long as we choose a renormalization scale $\mu\simeq v_\text{H}$ such that a perturbative analysis is valid close to the vacuum, the couplings $\lambda_\text{H}$ and $\lambda_\text{HS}$ are related, as demonstrated in Fig.~\ref{fig:why_cannot_mu=246gev}.
The ratio of the Higgs boson mass and the VEV, $m_\text{h}/v_\text{H}$, varies along each line shown in the figure.
Thus, if we fix $m_\text{h}=126$~GeV and $v_\text{H}=246$~GeV, the values of $\lambda_\text{H}$ and $\lambda_\text{HS}$ are fixed.
The values of $\lambda_\text{H}$ and $\lambda_\text{HS}$ for various $N$ are shown in Tab.~\ref{tab:lambdaH+lambdaP}.
These values are consistent with our order estimate Eq.~(\ref{eq:MN_order-count_01}).
If we use RG-improved potentials, $\lambda_\text{H}$ and $\lambda_\text{HS}$ are no longer fixed, since they depend on the value of $\lambda_\text{S}$.
\begin{table}[tbp]
\begin{center}
\begin{tabular}{|c||c|c|c|c|c|c|}
\hline
$N$	&$1$	&$2$	&$3$	&$4$	&$5$	&$12$
\\
\hline
$\lambda_\text{H}(v_\text{H})$	&$-0.11$	&$-0.055$&$-0.025$&$-0.0045$&$0.012$&$0.075$
\\
\hline
$\lambda_\text{HS}(v_\text{H})$	&$4.8$		&$3.4$&$2.8$&$2.4$&$2.1$&$1.4$
\\
\hline
\end{tabular}
\end{center}
\caption{\small 
Values of $\lambda_\text{H}$ and $\lambda_\text{HS}$ at $\mu= v_\text{H}$ fixed by $m_\text{h}=126$~GeV and $v_\text{H}=246$~GeV, in the case $V_\text{eff}=V_\text{LO}+V_\text{NLO}$.
}
\label{tab:lambdaH+lambdaP}
\end{table}

Let us present results of our analysis for the case $N=1$.
\begin{table}[tbp]
\begin{center}
\begin{tabular}{|c||c|c|c|}
\hline
&\multicolumn{3}{c|}{$N=1$}
\\
\hline
	&(I)	&(II)	&(III)
\\
\hline
\hline
$\mu$	&\multicolumn{3}{c|}{$v_\text{H}=246\,[\text{GeV}]$}
\\
\hline
$y_\text{t}(v_\text{H})$	&\multicolumn{3}{c|}{$0.919$}
\\
\hline
$g(v_\text{H})$	&\multicolumn{3}{c|}{$0.644$}
\\
\hline
$g'(v_\text{H})$	&\multicolumn{3}{c|}{$0.359$}
\\
\hline
$\lambda_\text{H}(v_\text{H})$	&$-0.11$	&$-0.059$	&$-0.082$
\\
\hline
$\lambda_\text{HS}(v_\text{H})$	&$4.8$	&$4.5$	&$4.3$
\\
\hline
$\lambda_\text{S}(v_\text{H})$	&$0.10$	&$0.10$	&$0.10$
\\
\hline
\hline
$v_\text{H}[\text{GeV}]$	&\multicolumn{3}{c|}{$246$}
\\
\hline
$m_\text{h}[\text{GeV}]$	&\multicolumn{3}{c|}{$126$}
\\
\hline
$\braket{\varphi}[\text{GeV}]$	&$0$	&$0$	&$0$
\\
\hline
$m_\text{s}[\text{GeV}]$	&$556$	&$527$	&$524$
\\
\hline
$\sin\theta_\text{mix}$	&$0$	&$0$	&$0$
\\
\hline
Landau pole [TeV]	&$3.5$	&$4.1$	&$4.7$
\\
\hline
\end{tabular}
\end{center}
\caption{\small 
Predictions of our model, together with some representative input parameters, for $N=1$ in three different approximations of the effective potentials: (I) $V_\text{LO}+V_\text{NLO}$, (II) $V_\text{eff}^\text{(LL)}$, and (III) $V_\text{eff}^{(\mbox{\scriptsize imp-NLO})}$.
In each case, a parameter set for $(\lambda_\text{H},\,\lambda_\text{HS},\,\lambda_\text{S})$ is chosen such that the mass and VEV of the Higgs boson are reproduced.
}
\label{tab:result_01}
\end{table}
Tab.~\ref{tab:result_01} shows our predictions, together with representative input parameters.
We compare the three different approximations of the effective potential: (I) $V_\text{LO}+V_\text{NLO}$, (II) $V_\text{eff}^\text{(LL)}$, and (III) $V_\text{eff}^{(\mbox{\scriptsize imp-NLO})}$.
Here, $V_\text{eff}^\text{(LL)}$ and $V_\text{eff}^{(\mbox{\scriptsize imp-NLO})}$ are defined similarly to Eqs.~(\ref{eq:VLL}) and (\ref{eq:effpot_ms-bar_mod_improved}) from $V_\text{LO}$ and  $V_\text{LO}+V_\text{NLO}$ with the scale choice $t=\ln(\sqrt{{\phi}^2+{\varphi}^2}/v_\text{H})$, respectively.
As mentioned, in the case (I), $\lambda_\text{H}$ and $\lambda_\text{HS}$ are fixed by $m_\text{h}$ and $v_\text{H}$, while $\lambda_\text{S}$ can be taken as a free input parameter.
Reflecting this feature, even in the cases (II) and (III), the values of $\lambda_\text{H}$ and $\lambda_\text{HS}$ are tightly constrained, while $\lambda_\text{S}$ can be taken fairly freely.
For this reason, we take $\lambda_\text{S}$ as the input parameter for all three cases.
In the table we take $\lambda_\text{S}=0.10$ as an example.
In all cases there exist $\lambda_\text{H}$ and $\lambda_\text{HS}$ which reproduce the mass and VEV of the Higgs boson. 
Furthermore, in accord with the argument in Sec.~\ref{s3.1}, the mass of the singlet scalar is predicted to be several times larger than the Higgs boson mass in each case.
We find that the predicted masses of the singlet scalar are consistent with each other within about 5\% accuracy.
The differences may be taken as a reference for the stability of our predictions.
An undesirable feature is that the locations of the Landau pole are close and in the several TeV region.
This originates from the large value of $\lambda_\text{HS}$ at the electroweak scale $\mu \simeq v_\text{H}$ (see Sec.~\ref{s3.1}).

Dependences of the predictions on $\lambda_\text{S}$ is as follows.
If we raise the value of $\lambda_\text{S}$, the locations of the Landau pole are even lowered, since the couplings $\lambda_\text{H}$, $\lambda_\text{HS}$, $\lambda_\text{S}$ increase with the renormalization scale as they influence each other.
The mass of the singlet scalars are barely dependent on
$\lambda_\text{S}$, since only the fourth derivative in the $\varphi$ direction 
of the effective potential is affected at LO.
The values of $\lambda_\text{H}$ and $\lambda_\text{HS}$ do not change very much with $\lambda_\text{S}$.
For instance, if we take $\lambda_\text{S}=0.5$ and $1.0$, the Landau pole appears at $3.2$TeV and $2.8$TeV respectively, while the mass of the singlet scalar changes little.

\begin{figure}[tbp]
	\begin{center}
		\begin{tabular}{cc}
		\hspace*{-0.3cm}
		\includegraphics[width=8cm]{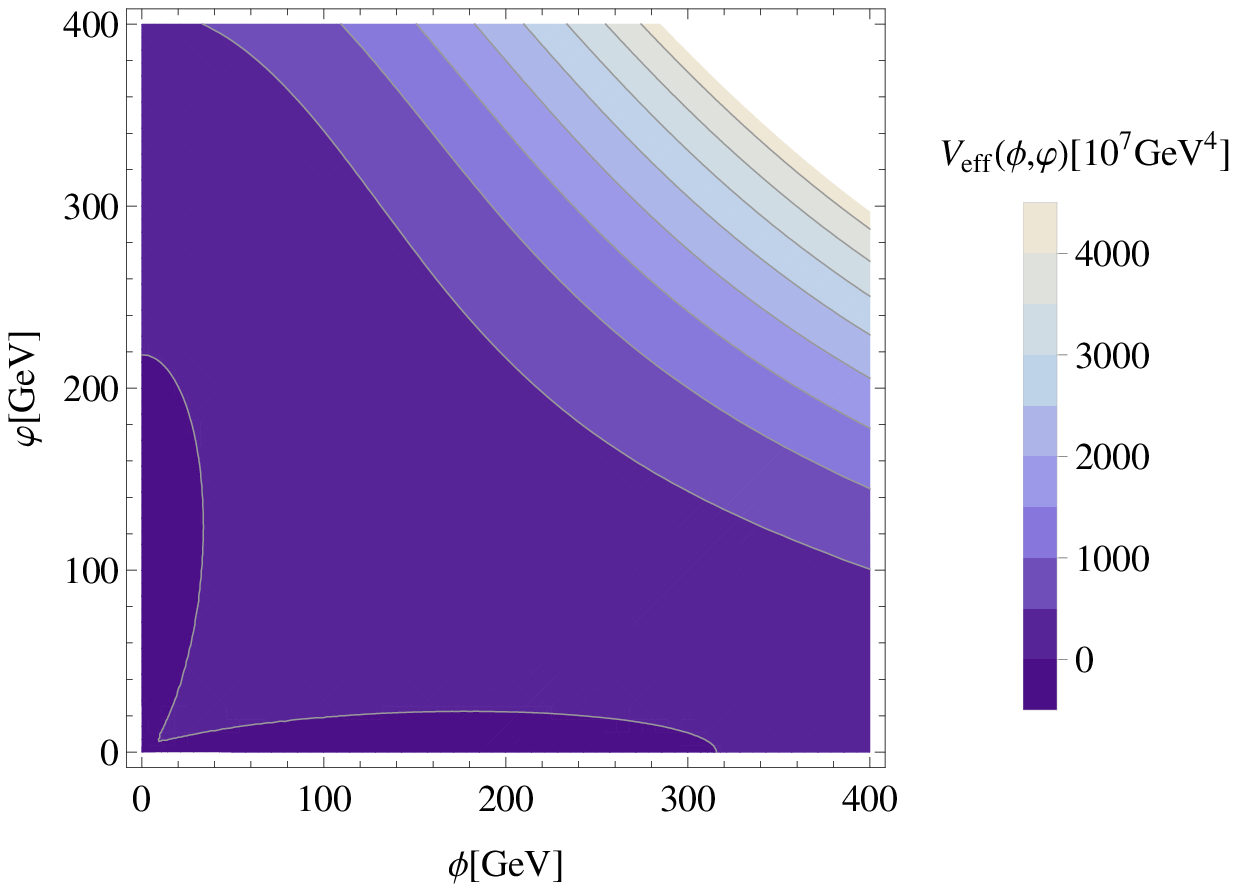}
		&
		\includegraphics[width=7.5cm]{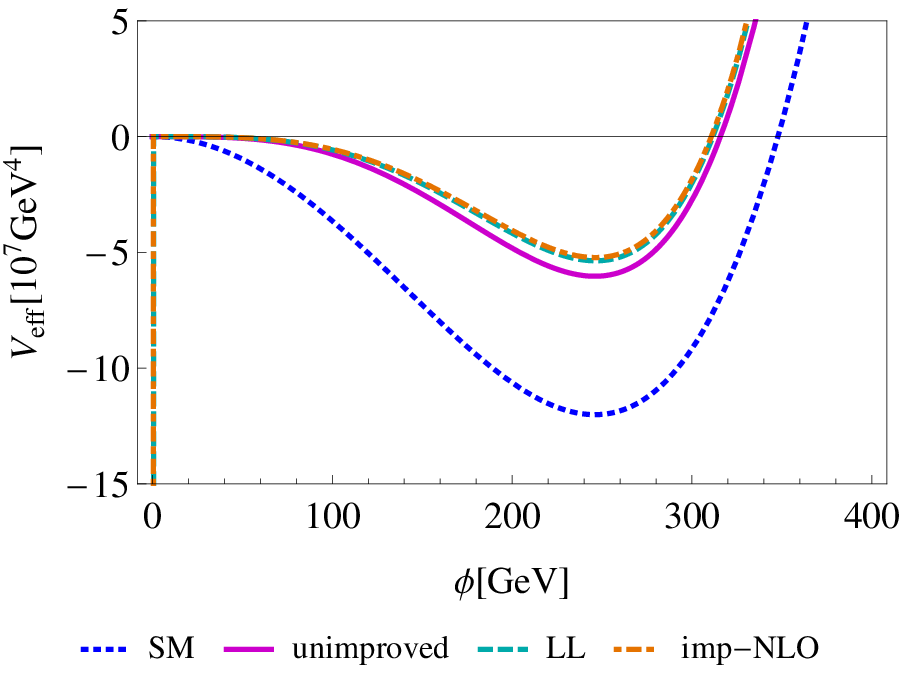}
		\\
		(a)&(b)
		\end{tabular}
	\end{center}
\caption{\small  (a)
Contour plot of the unimproved effective potential $V_\text{LO}+V_\text{NLO}$ .
(b) Comparison of the effective potentials on the $\phi$-axis ($\varphi=0$), in the approximations (I) $V_\text{LO}+V_\text{NLO}$, (II) $V_\text{eff}^\text{(LL)}$, and (III) $V_\text{eff}^{(\mbox{\scriptsize imp-NLO})}$.
The effective potential for the SM is also shown.
}
\label{fig:ui_fig}
\end{figure}

To see the  shape of the effective potential, we show in Fig.~\ref{fig:ui_fig} (a) the contour plot of the potential in case (I) of Tab.~\ref{tab:result_01}.
Since the potential is symmetric under $\phi\to -\phi$ or $\varphi\to - \varphi$, we show only the upper-right part of the configuration space.
As expected there is a global minimum along the $\phi$-axis.
There is also a shallower local minimum along the $\varphi$-axis, which is generated by the CW mechanism of a competition between $\lambda_\text{S} \varphi^4$ and ${\lambda_\text{HS}}^2\varphi^4\ln(\lambda_\text{HS}\varphi^2/\mu^2)$ terms.\footnote{
This local minimum on the $\varphi$-axis is about ten times shallower than the global minimum.
It becomes even shallower if the value of $\lambda_\text{S}$ is larger. 
We will not be concerned about the  minimum on the $\varphi$-axis in our analysis as long as it is not a global minimum.
}
In cases (II) and (III) of Tab.~\ref{tab:result_01}, we obtain qualitatively similar contour plots.

In Fig.~\ref{fig:ui_fig} (b) we show the effective potentials on the $\phi$-axis (i.e., for $\varphi=0$) for all three cases, together with the effective potential of the SM.
We see that differences between the three approximations are small.
In particular, the difference between the LL approximation (II) and the improved-NLO approximation (III) is hardly visible.
These features show good stability of our predictions in the vicinity of the vacuum, and that they are within the validity range of perturbation theory.

There is a significant difference between our effective potential and that of the SM.
In comparison to the SM, the minimum of our potential is shallow, although the values of $v_\text{H}$ and $m_\text{h}$ are common.
The clear difference of the potential shapes indicates that the higher derivatives of the potentials at the minimum are appreciably different.
Namely, we anticipate that the Higgs self-interactions of our model are appreciably different from those of the SM.

\begin{figure}[tbp]
	\begin{center}
		\begin{tabular}{cc}
		\hspace*{-0.3cm}
		\includegraphics[width=8cm]{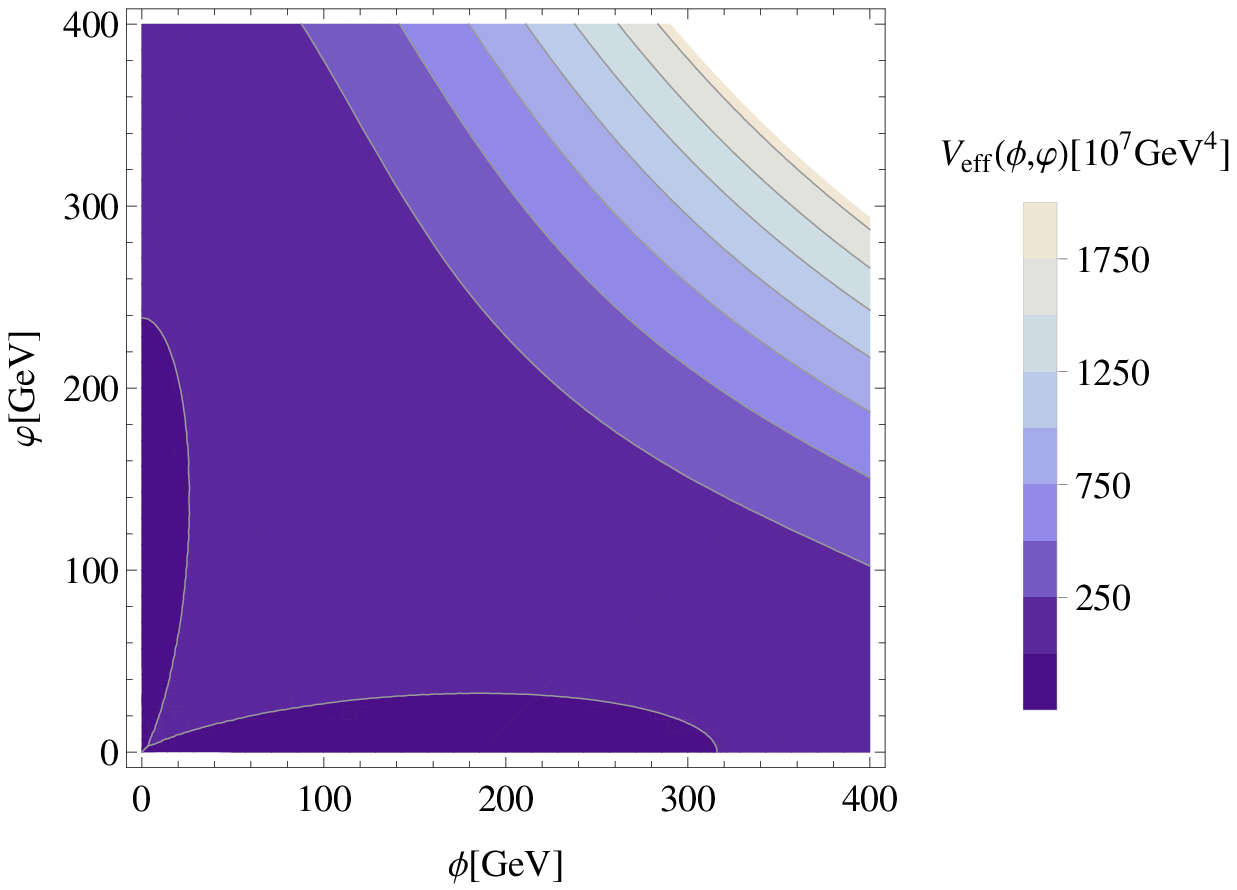}
		&
		\includegraphics[width=7.5cm]{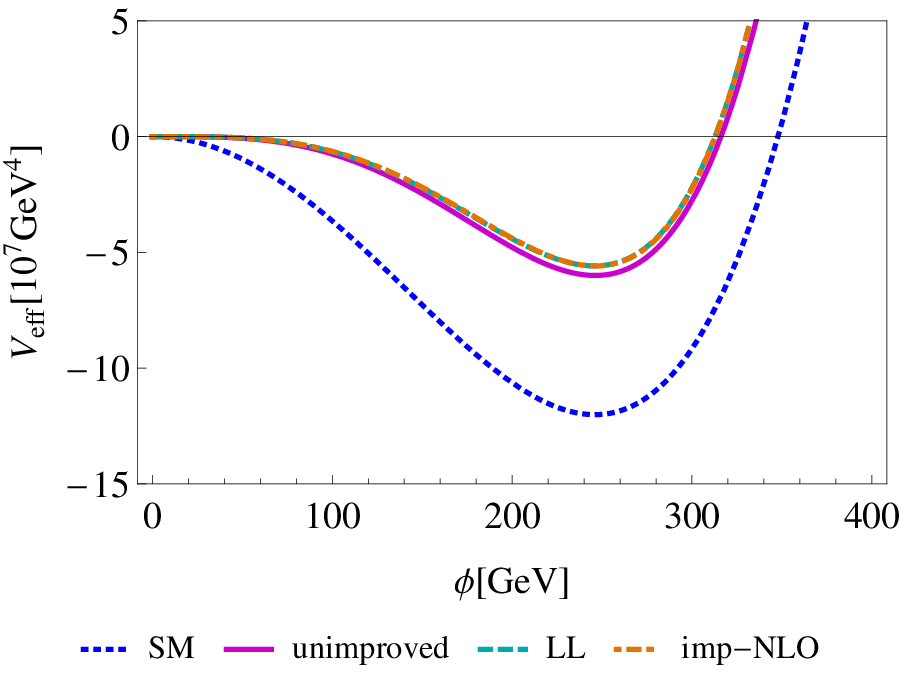}
		\\
		(a)&(b)\vspace*{10mm}\\
		\includegraphics[width=8cm]{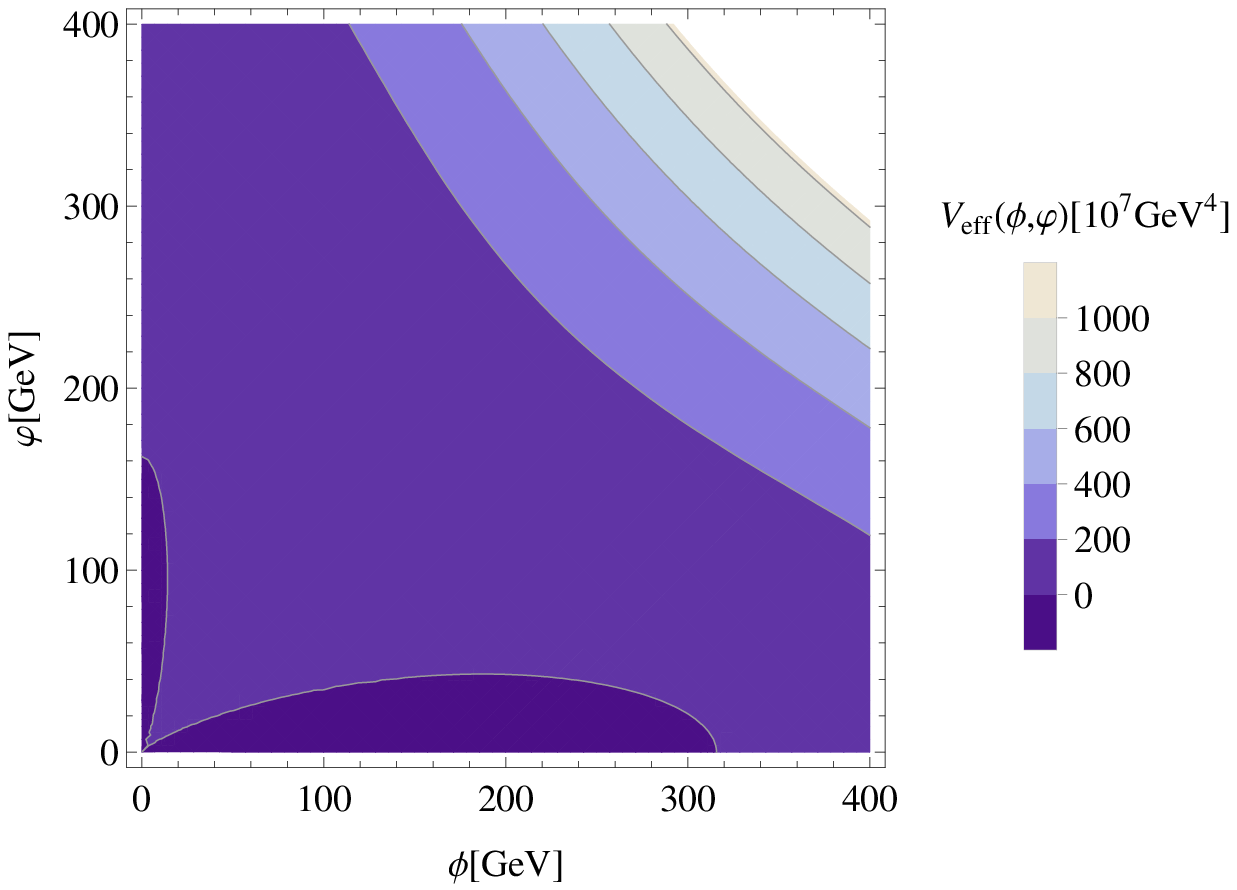}
		&
		\includegraphics[width=7.5cm]{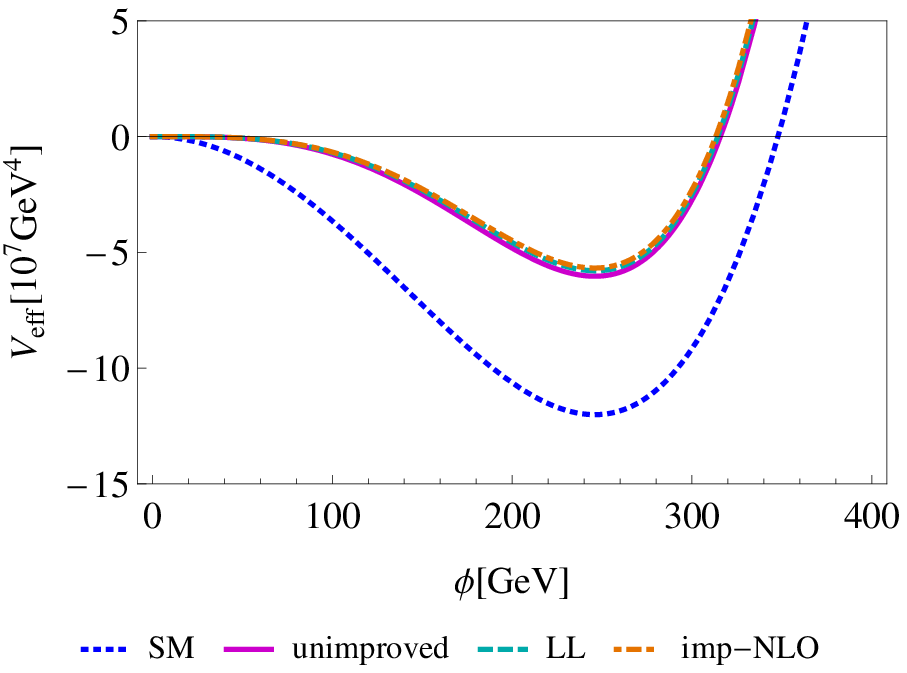}
		\\
		(c)&(d)\\
		\end{tabular}
	\end{center}
\caption{\small  Same as Figs.~\ref{fig:ui_fig} but for (a)(b) $N=4$, and (c)(d) $N=12$, corresponding to the parameters of Tab.~\ref{tab:N_more_than_1}.
}
\label{fig:pot-largeN}
\end{figure}

Next we examine the cases $N>1$.
In general, we expect that the location of the Landau pole is raised as compared to the $N=1$ case.
This is because the required value of $\lambda_\text{HS}$ to overwhelm the top-loop contribution decreases with $N$ as $1/\sqrt{N}$, see Sec.~\ref{s3.1}.
One may confirm this tendency in Tab.~\ref{tab:lambdaH+lambdaP}.
We show the cases $N=4$ and $N=12$ in Tab.~\ref{tab:N_more_than_1} with the input $\lambda_\text{S}=0.10$ at $\mu=v_\text{H}$. 
As expected the positions of the Landau pole are raised up to order a few tens TeV for these $N$.
The corresponding shapes of the effective potentials are displayed in Figs.~\ref{fig:pot-largeN}.
\begin{table}[tbp]
\begin{center}
\begin{tabular}{|c||c|c|c||c|c|c|}
\hline
&\multicolumn{3}{c|}{$N=4$}
&\multicolumn{3}{c|}{$N=12$}
\\
\hline
	&(I)	&(II)	&(III)
	&(I)	&(II)	&(III)
\\
\hline
\hline
$\lambda_\text{H}(v_\text{H})$	
&$-0.0045$	&$-0.061$	&$-0.0005$
&$0.075$	&$-0.063$	&$0.082$
\\
\hline
$\lambda_\text{HS}(v_\text{H})$	
&$2.4$	&$2.3$	&$2.4$
&$1.4$	&$1.4$	&$1.4$
\\
\hline
$\lambda_\text{S}(v_\text{H})$	
&$0.10$	&$0.10$	&$0.10$
&$0.10$	&$0.10$	&$0.10$
\\
\hline
\hline
$\braket{\varphi}[\text{GeV}]$	
&$0$	&$0$	&$0$
&$0$	&$0$	&$0$
\\
\hline
$m_\text{s}[\text{GeV}]$	
&$378$	&$378$	&$375$
&$285$	&$293$	&$286$
\\
\hline
$\sin\theta_\text{mix}$	
&$0$	&$0$	&$0$
&$0$	&$0$	&$0$
\\
\hline
Landau pole [TeV]	
&$16$	&$19$	&$17$
&$28$	&$37$	&$26$
\\
\hline
\end{tabular}
\end{center}
\caption{\small 
Predictions of our model for $N=4$ and $N=12$ in three different approximations of the effective potentials:
(I) $V_\text{LO}+V_\text{NLO}$, (II) $V_\text{eff}^\text{(LL)}$, and (III) $V_\text{eff}^{(\mbox{\scriptsize imp-NLO})}$. 
In each case, a parameter set for $(\lambda_\text{H},\,\lambda_\text{HS},\,\lambda_\text{S})$ is chosen such that the mass and VEV of the Higgs boson are reproduced.
}
\label{tab:N_more_than_1}
\end{table}

\begin{figure}[tbp]
	\begin{center}
\begin{tabular}{cc}
		\includegraphics[width=7cm]{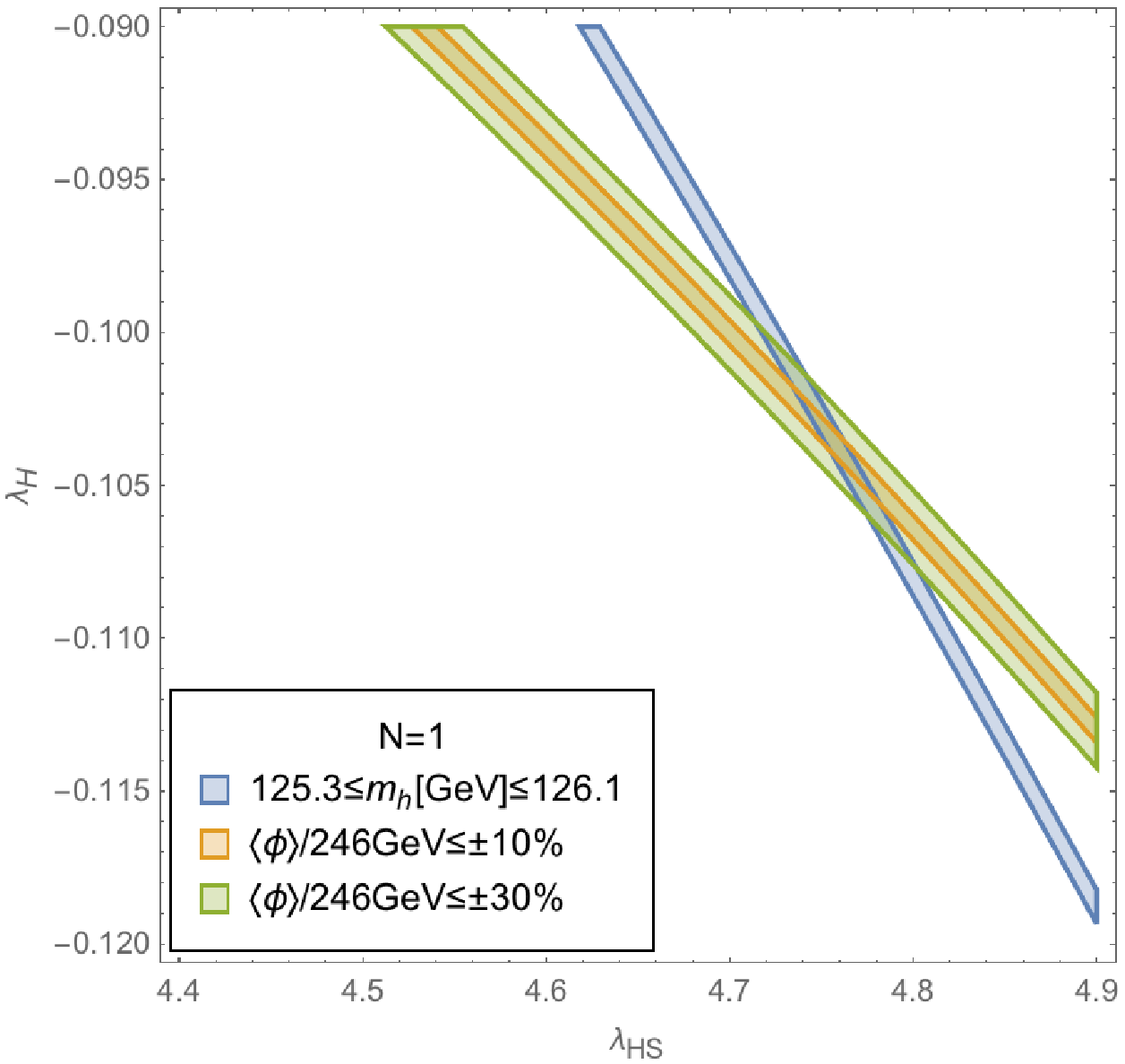}
		&\includegraphics[width=7cm]{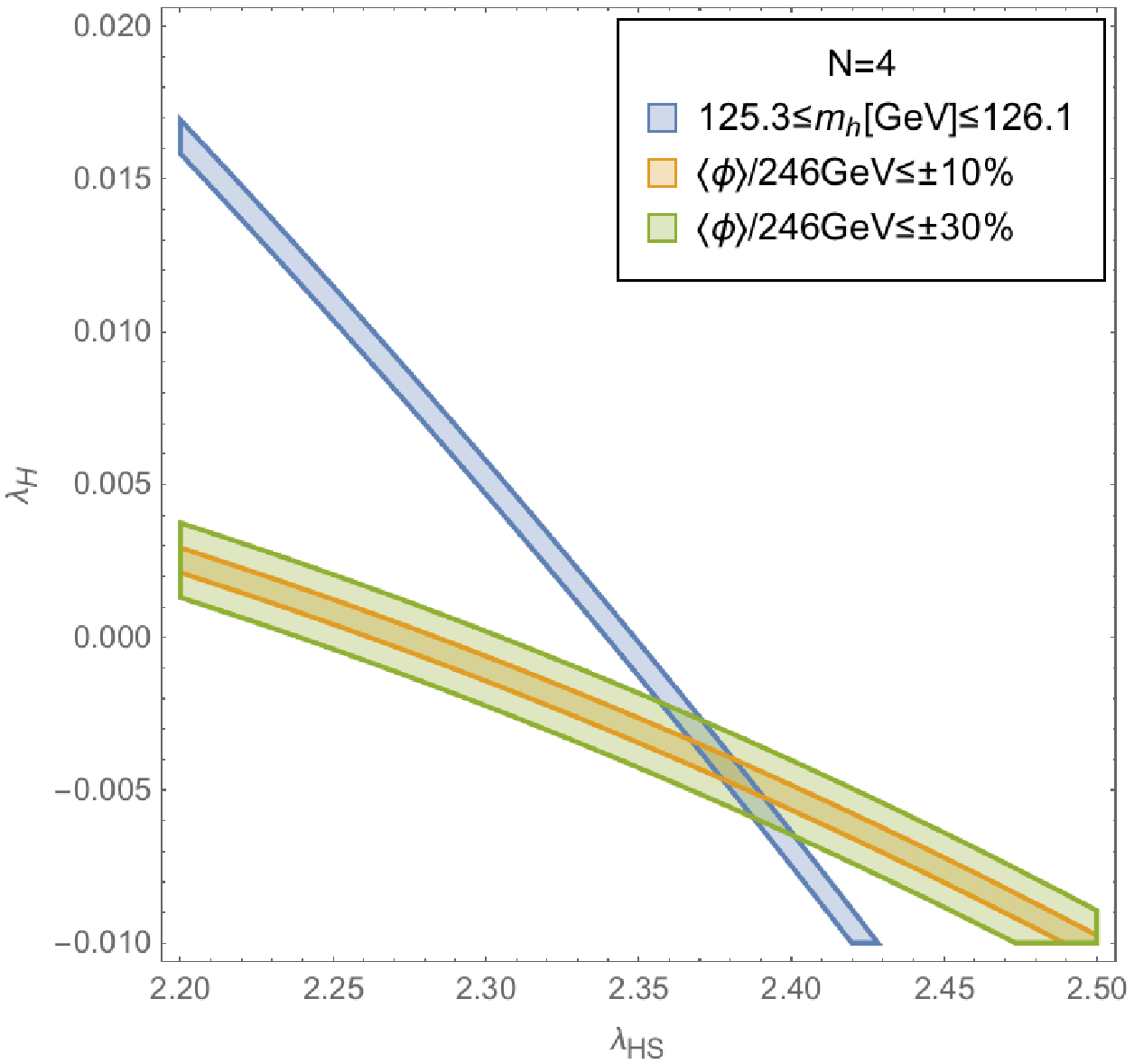}
\\
	(a)	&(b)
\\
\\
	\multicolumn{2}{c}{\includegraphics[width=7cm]{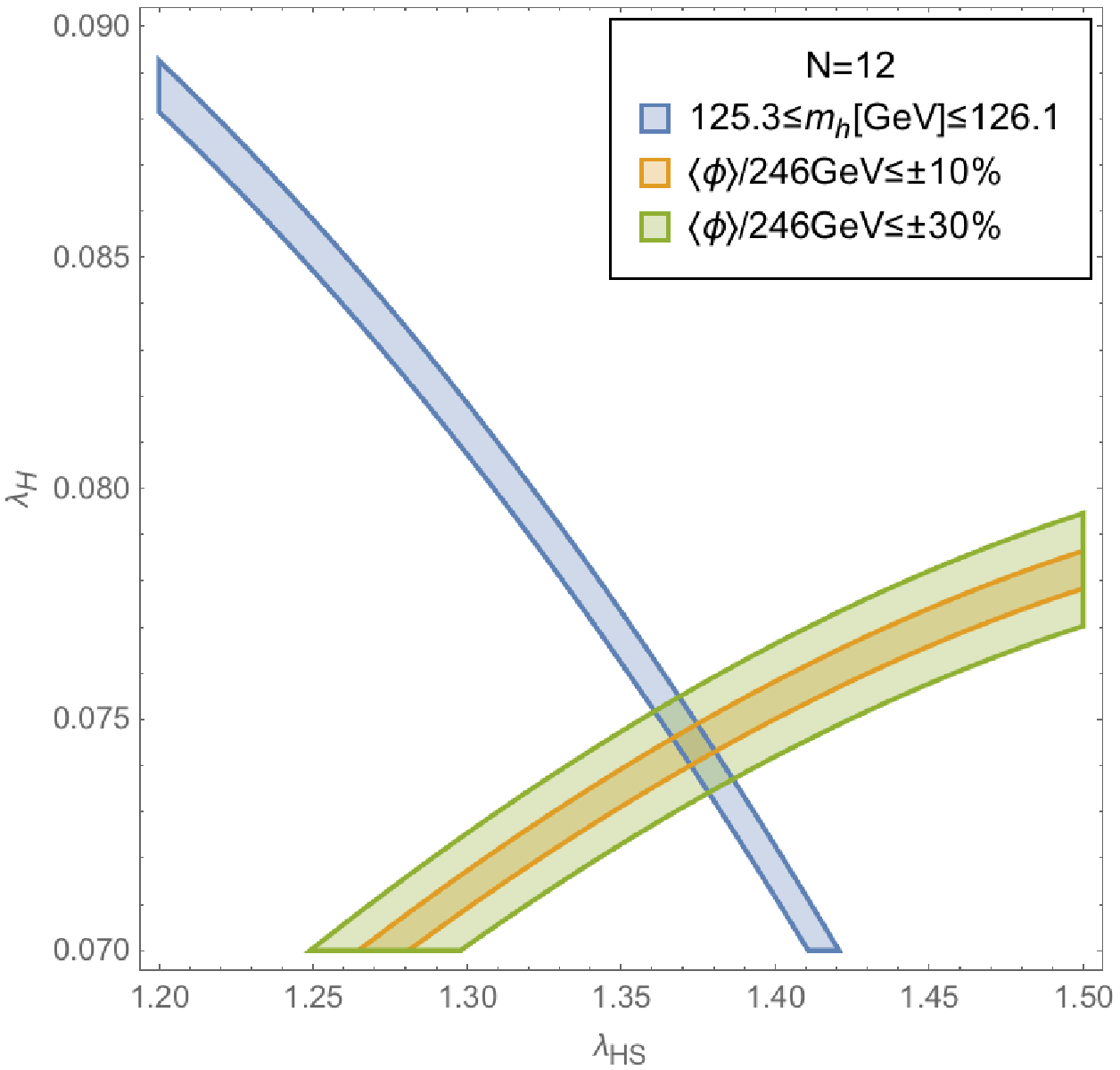}}
\\
	\multicolumn{2}{c}{(c)}
\end{tabular}
	\end{center}
\caption{\small 
Plots for
phenomenologically favoured region of 
the portal coupling $\lambda_\text{HS}$ and the Higgs quartic coupling
$\lambda_\text{H}$ for
(a) $N=1$, (b) $4$ and (c) $12$, 
corresponding to the experimental data \cite{Agashe:2014kda}. 
$\lambda_\text{S}$ is set as 0.1, although the results are
fairly insensitive to $\lambda_\text{S}$.
\label{fig:rpPheno}
}
\end{figure}

In Fig.~\ref{fig:rpPheno} 
are shown the plots of the phenomenologically favored region for 
the portal coupling $\lambda_\text{HS}$ and the Higgs quartic coupling
$\lambda_\text{H}$.
We show the results for the case (I), which are independent of 
$\lambda_\text{S}$.
The other parameters are fixed to the SM values.
These figures show how the parameters of the model under
consideration are constrained
in the current status.
As we discussed, currently $\lambda_\text{S}$ is barely constrained.

\subsection{Interactions among physical scalar particles}

\subsubsection{Our results}
Let us expand the effective potential about the vacuum after rewriting $\phi^2\to H^{\dagger}H$, $\varphi^2 \to \vec{S}\cdot\vec{S}$ and 
\begin{align}
	H
=&	\frac{1}{\sqrt{2}}
	\left(
	\begin{array}{c}
		0\\
		v_\text{H}+h
	\end{array}
	\right)\,,
\nonumber\\
	\vec{S}
=&	\,(v_\text{S}+s_1,\,s_2,\,\cdots,\,s_N)^\text{T}
\nonumber\\
=&	\,(s_1,\,s_2,\,\cdots,\,s_N)^\text{T}
\equiv
	\,\vec{s}\,.
\label{eq:field_def_02}
\end{align}
Then the effective potential takes a form
\begin{align}
	V_\text{eff}
=&	\, 
	\text{const.}
	+\frac{1}{2}m_\text{h}^2 h^2
	+\frac{1}{2}m_\text{s}^2 \,\vec{s}\cdot\vec{s}
	+\frac{\lambda_\text{hhh}}{3!}\,v_\text{H}h^3
	+\frac{\lambda_\text{hhhh}}{4!}h^4
\nonumber\\
&	+\frac{\lambda_\text{hss}}{2}\,v_\text{H}\,h\,\vec{s}\cdot\vec{s}
	+\frac{\lambda_\text{hhss}}{4}h^2\,\vec{s}\cdot\vec{s}
+\frac{\lambda_\text{ssss}}{4!}\,(\vec{s}\cdot\vec{s})^2
+ \dots ,
\label{expandVeff}
\end{align}
where only up to dimension-four interactions are shown explicitly.

\begin{table}[tbp]
\begin{center}
\begin{tabular}{|c||c|c|c||c|c|c||c|c|c|}
\hline
&\multicolumn{3}{c||}{$N=1$}
&\multicolumn{3}{c||}{$N=4$}
&\multicolumn{3}{c|}{$N=12$}
\\
\hline
	&(I)	&(II)	&(III)
	&(I)	&(II)	&(III)
	&(I)	&(II)	&(III)
\\
\hline
\hline
$\lambda_\text{hhh}/\lambda^\text{(SM)}_\text{hhh}$	
&$1.7$	&$1.8$	&$1.8$
&$1.7$	&$1.7$	&$1.7$
&$1.7$	&$1.6$	&$1.7$
\\
\hline
$\lambda_\text{hhhh}/\lambda^\text{(SM)}_\text{hhhh}$	
&$3.7$	&$4.3$	&$4.5$
&$3.7$	&$3.2$	&$3.4$
&$3.7$	&$2.8$	&$3.1$
\\
\hline
$\lambda_\text{hss}$	
&$11.4$	&$10.2$	&$10.2$
&$5.02$	&$5.02$	&$4.96$
&$2.80$	&$2.95$	&$2.83$
\\
\hline
$\lambda_\text{hhss}$	
&$14$	&$13$	&$13$
&$5.6$	&$5.7$	&$5.7$
&$3.0$	&$3.2$	&$3.1$
\\
\hline
$\lambda_\text{ssss}$	
&$-$	&$6.5$	&$-$
&$-$	&$1.9$	&$-$
&$-$	&$0.9$	&$-$
\\
\hline
\end{tabular}
\end{center}
\caption{\small 
Coupling constants among the scalar particles, corresponding to the parameters of Tabs.~\ref{tab:result_01} and \ref{tab:N_more_than_1}.
The coupling constants are defined in Eq.~(\ref{expandVeff}).
}
\label{tab:result_02}
\end{table}

We show the values of the coupling constants among the scalar particles in Tab.~\ref{tab:result_02}.
They correspond to the parameters of Tabs.~\ref{tab:result_01} and \ref{tab:N_more_than_1}.
In the case $N=1$, the triple self-coupling of the Higgs boson
$\lambda_\text{hhh}$ turns out to be larger than the SM value by a factor 1.7--1.8, while the quartic self-coupling 
$\lambda_\text{hhhh}$ is larger by a factor 3.7--4.5.
The range of each value shows the level of accuracy of our prediction.
(In general higher derivatives of the potential have larger uncertainties.)
These values are barely dependent on the input value for $\lambda_\text{S}$, since the effective potential in case (I) is independent of $\lambda_\text{S}$ on the $\phi$-axis, and the dependence is weak in the cases (II) and (III).

For $N=1$, the coupling constants involving the singlet scalars are large and of order ten.
They become even larger if the input value for $\lambda_\text{S}$ is taken to be larger.
One may wonder if perturbation theory is valid with such large coupling constants.
We remind the reader that the coupling constants in the original Lagrangian are not so large, and that our predictions have been tested to be within the perturbative regime in the vicinity of the scale of the vacuum.
These large couplings are considered to be a typical feature of the 
present model 
and need to be tested by future experiments.

For larger $N$, the Higgs triple self-coupling is barely dependent on $N$, while the quartic coupling decreases slightly with $N$.
The $N$ dependences of the couplings involving the singlet scalars are more evident.
Generally we obtain smaller couplings for larger $N$.
This originates from high sensitivities of these couplings on $\lambda_\text{HS}$, which reduces with $N$ as $\sim 1/\sqrt{N}$, and can be regarded as a characteristic feature of our potential.

In the cases (I) and (III) the values of $\lambda_\text{ssss}$ are not shown in the table, for the following reason.
If we compute the fourth derivative of $V_\text{NLO}$ with respect to $\vec{S}$ and set $h=|\vec{s}|=0$, the contribution of the NG modes of $H$ diverges.
This does not happen in the case (II) since $V_\text{NLO}$ is not involved.
The divergence originates from an infra-red region, $k\sim 0$ in $\int d^4k/(k^2)^2$, and is an artifact of setting all the external momenta to zero.
In physical amplitudes, momenta flowing into the quartic vertex are almost always non-zero, which regularize the IR divergence in the loop integral.
Since the other couplings have similar values for the cases (I), (II), (III), we expect that the values of $\lambda_\text{ssss}$ for the case (II) 
would give reasonable estimates of the quartic vertex appropriate for physical amplitudes.

\begin{table}[tbp]
\begin{center}
\begin{tabular}{|c|c||c|c|c|c|c|}
\hline
\multicolumn{2}{|c||}{$N=1$ case}	&$\lambda_\text{hhh}$	&$\lambda_\text{hhhh}$	&$\lambda_\text{hss}$	&$\lambda_\text{hhss}$	&$\lambda_\text{ssss}$	
\\
\hline
\hline
\multicolumn{2}{|c||}{SM prediction}	&$0.78$	&$0.78$	&none	&none	&none	
\\
\hline
\hline
\multicolumn{2}{|c||}{GW's framework}	&$1.3$	&$2.9$	&$2\lambda_\text{HS}=9.6$	&$2\lambda_\text{HS}=9.6$	& $6\lambda_\text{S}=0.6$
\\
\hline
\hline
	&(I)	&$1.3$	&$2.9$	&$11.4$	&$13.8$	&-
\\
\cline{2-7}
our analysis	&(II)		&$1.4$	&$3.4$	&$10.2$	&$13.0$	&$6.5$
\\
\cline{2-7}
	&(III)	&$1.4$	&$3.6$	&$10.2$	&$13.5$	&-
\\
\hline
\end{tabular}
\vspace{3mm}\\
(a)
\vspace{5mm}\\

\begin{tabular}{|c|c||c|c|c|c|c|}
\hline
\multicolumn{2}{|c||}{$N=4$ case}	&$\lambda_\text{hhh}$	&$\lambda_\text{hhhh}$	&$\lambda_\text{hss}$	&$\lambda_\text{hhss}$	&$\lambda_\text{ssss}$	
\\
\hline
\hline
\multicolumn{2}{|c||}{SM prediction}	&$0.78$	&$0.78$	&none	&none	&none	
\\
\hline
\hline
\multicolumn{2}{|c||}{GW's framework}	&$1.3$	&$2.9$	&$2\lambda_\text{HS}=4.8$	&$2\lambda_\text{HS}=4.8$	& $6\lambda_\text{S}=0.6$
\\
\hline
\hline
	&(I)	&$1.3$	&$2.9$	&$5.0$	&$5.6$	&-
\\
\cline{2-7}
our analysis	&(II)		&$1.3$	&$2.5$	&$5.0$	&$5.7$	&$1.9$
\\
\cline{2-7}
	&(III)	&$1.4$	&$2.7$	&$5.0$	&$5.7$	&-
\\
\hline
\end{tabular}
\vspace{3mm}\\
(b)
\vspace{5mm}\\

\begin{tabular}{|c|c||c|c|c|c|c|}
\hline
\multicolumn{2}{|c||}{$N=12$ case}	&$\lambda_\text{hhh}$	&$\lambda_\text{hhhh}$	&$\lambda_\text{hss}$	&$\lambda_\text{hhss}$	&$\lambda_\text{ssss}$	
\\
\hline
\hline
\multicolumn{2}{|c||}{SM prediction}	&$0.78$	&$0.78$	&none	&none	&none	
\\
\hline
\hline
\multicolumn{2}{|c||}{GW's framework}	&$1.3$	&$2.9$	&$2\lambda_\text{HS}=2.8$	&$2\lambda_\text{HS}=2.8$	& $6\lambda_\text{S}=0.6$
\\
\hline
\hline
	&(I)	&$1.3$	&$2.9$	&$2.8$	&$3.0$	&-
\\
\cline{2-7}
our analysis	&(II)		&$1.3$	&$2.2$	&$3.0$	&$3.2$	&$0.92$
\\
\cline{2-7}
	&(III)	&$1.3$	&$2.4$	&$2.8$	&$3.1$	&-
\\
\hline
\end{tabular}
\vspace{3mm}\\
(c)
\vspace{5mm}\\

\end{center}
\caption{\small Comparisons between the results derived by our analysis and those derived by GW's method for $N=1$, $4$ and $12$ cases in (a), (b) and (c),
respectively. 
}
\label{tab:coupling_comparison}
\end{table}

\subsubsection{Comparison with results using Gildener-Weinberg's framework}

Let us compare our results with those using the conventional
GW framework.

We show in Tab.~\ref{tab:coupling_comparison}(a)-(c)
comparisons of the couplings among the physical Higgs boson and singlet 
scalar(s),
including their self-couplings, as defined in eq.~(\ref{expandVeff}).
The effective potential in the GW framework coincides with our
effective potential after setting $\varphi=0$.
Hence, the Higgs self-couplings are the same in both analyses.
[We list the values for case (I) of our analysis as the
corresponding values in the GW framework.]

In conventional analyses using the GW framework,
the couplings involving the singlet scalars are derived
using the tree-level interactions (by substituting the
VEV of the Higgs boson appropriately).
Therefore, loop-induced effects are not included.
In our case, since the portal coupling $\lambda_\text{HS}$ is large,
its loop-induced effects can be large.
These effects are most enhanced for $N=1$, where $\lambda_\text{HS}$ is
the largest; see Tab.~\ref{tab:coupling_comparison}(a).
In particular
there is a large enhancement of the singlet quartic self-coupling
$\lambda_\text{ssss}$, if $\lambda_\text{S}$ is small
and the effect of $\lambda_\text{HS}$-induced loop contribution
is much larger.
(We list the case $\lambda_\text{S}=0.1$ in the table.)
Note that, although
$\lambda_\text{S}$ is practically a free parameter,
a smaller value is preferred with regard to the Landau pole.
For completeness, we show the $\lambda_\text{S}$ dependences
of our predictions in Tab.~\ref{tab1}.
We see that only $\lambda_\text{ssss}$ depends considerably on
$\lambda_\text{S}$, and the dependence is roughly consistent with
that of $6\lambda_\text{S}+\text{const}.$, as anticipated.

In view of the large portal coupling, certainly it is
sensible to include the loop-induced effects in computing 
the interactions involving the
singlet scalars.
In this case, we need to set up a formulation with proper
account of order counting, as described in 
the previous section.
It inevitably requires departure from the analysis
in a one-dimensional subspace, i.e., the GW formulation.

The differences between our results and those of
the GW framework tend to decrease for larger $N$,
since the portal coupling becomes smaller.
This tendency can be confirmed in 
Tabs.~\ref{tab:coupling_comparison}(b)(c).

\begin{table}
\begin{center}
\begin{tabular}{|c||c|c|c|c|c|}
\hline
$\lambda_\text{S}$			&0.1	&0.3	&0.5	&1.0	&2.0\\
\hline
\hline
$\lambda_\text{hss}$ (I)	&11.4	&11.5	&11.5	&11.7	&11.9\\
\hline
$\lambda_\text{hss}$ (II)	&10.2	&10.2	&10.3	&10.3	&10.4\\
\hline
$\lambda_\text{hss}$ (III)	&10.2	&10.3	&10.3	&10.4	&10.4\\
\hline
$\lambda_\text{ssss}$ (II)	&6.53	&7.91	&9.32	&12.8	&19.7\\
\hline
\end{tabular}
\end{center}
\caption{\label{tab1}
$\lambda_\text{S}$ dependences
of the couplings involving the singlet scalar bosons.
}
\end{table}

\begin{figure}[tbp]
	\begin{center}
\begin{tabular}{cc}
		\includegraphics[width=7cm]{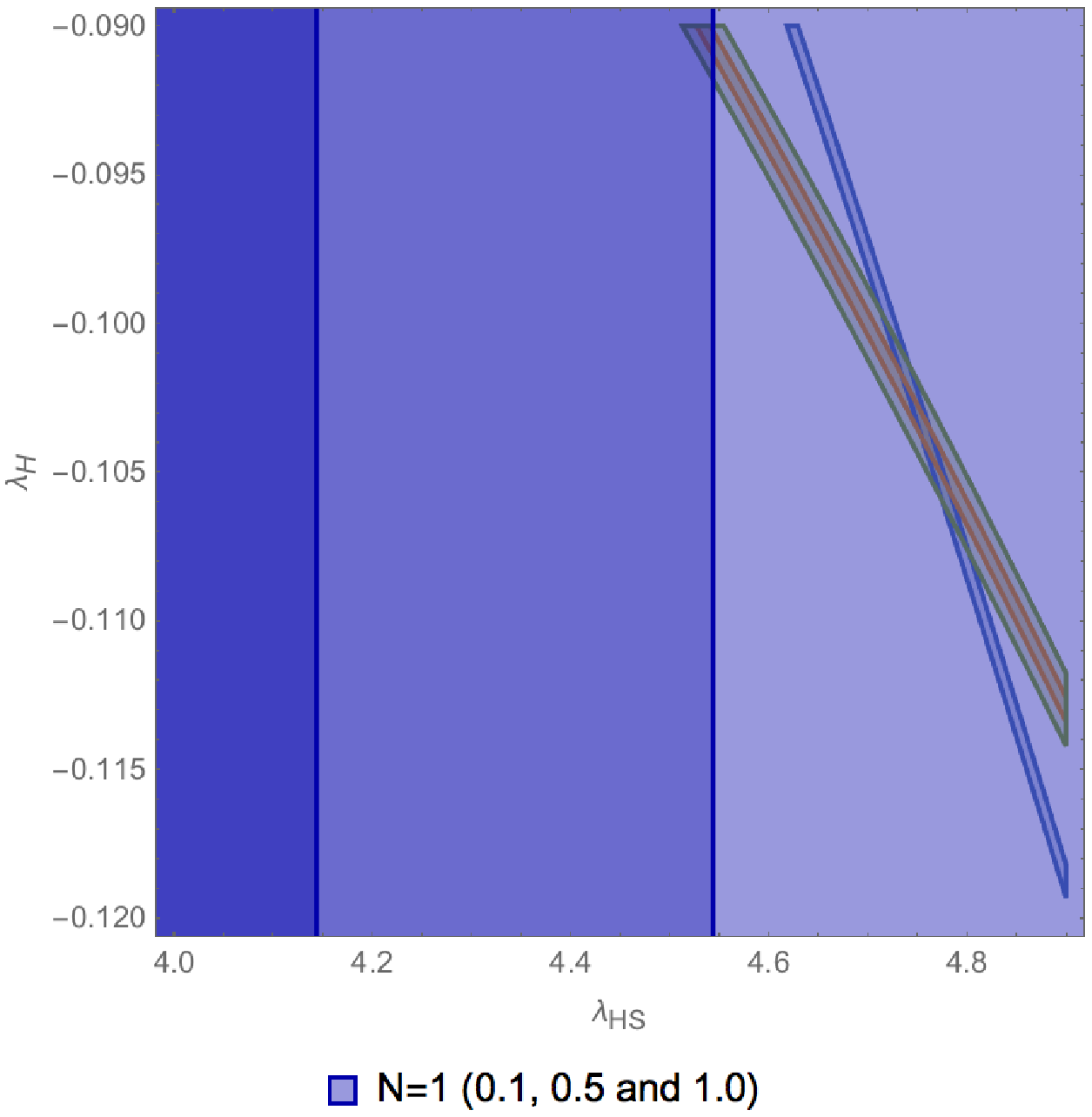}
		&\includegraphics[width=7cm]{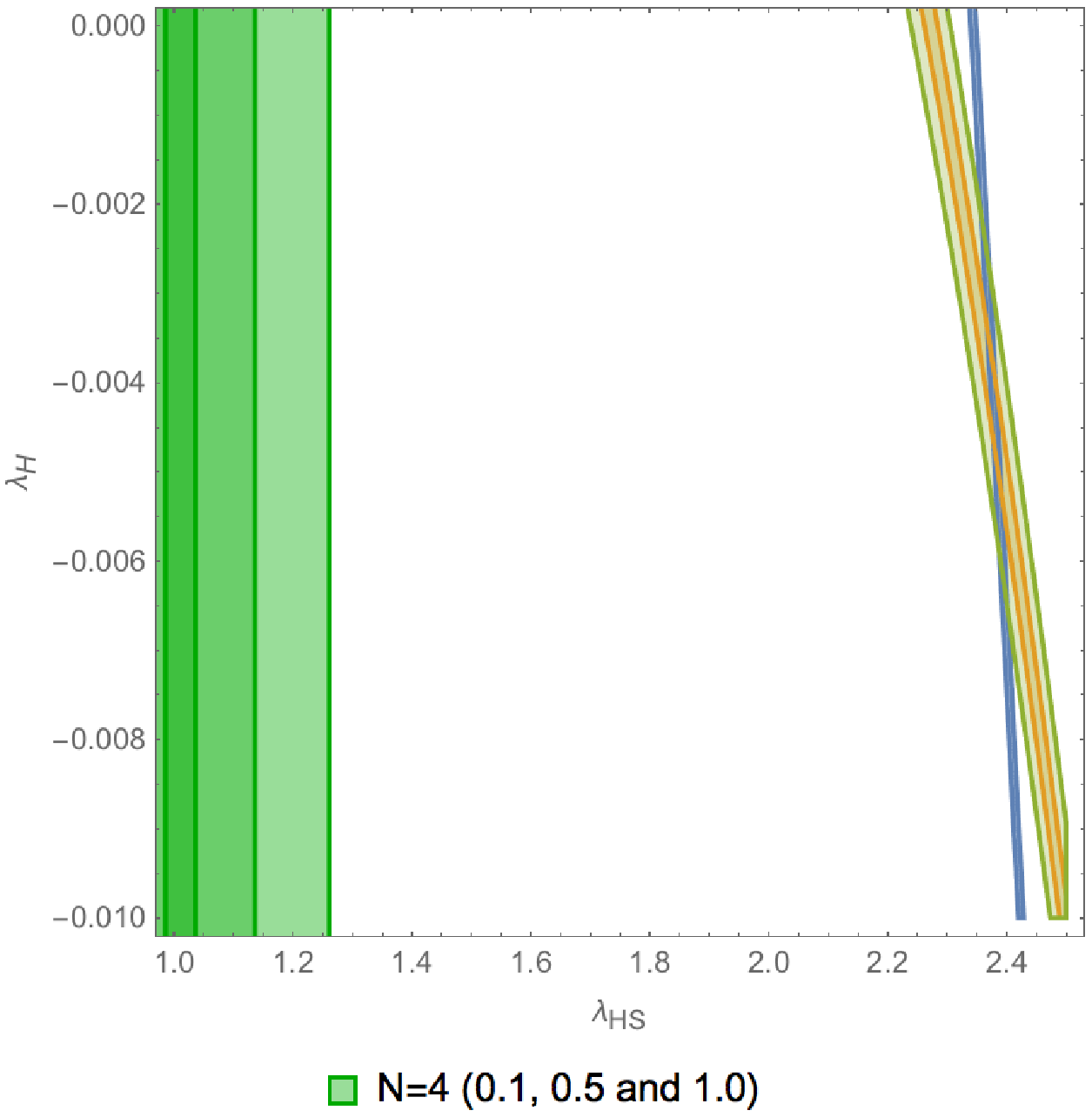}
\\
	(a)	&(b)
\\
\multicolumn{2}{c}{\includegraphics[width=7cm]{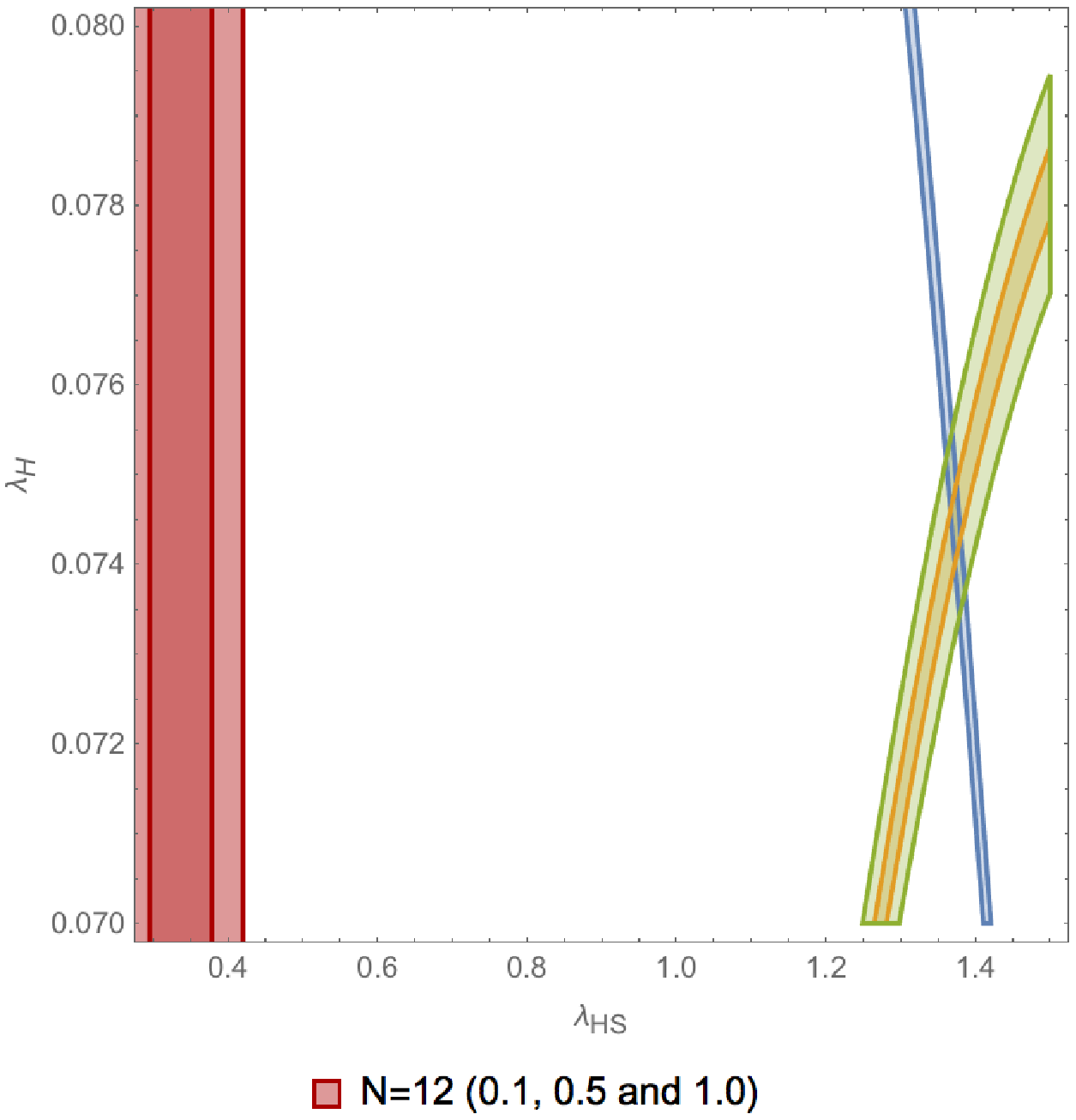}}
\\
\multicolumn{2}{c}{(c)}
\end{tabular}
	\end{center}
\caption{\small Favoured regions of Veltman's condition for the 
Higgs mass for $N=1$ [(a)], $N=4$ [(b)] and $N=12$ [(c)].
We combine the plot in each case with
the respective plot of Figs.~\ref{fig:rpPheno}. 
The regions in the
{\it darkest}, {\it darker} and {\it light} colors correspond to
the magnitudes of the coefficient of the quadratic divergence 
[given by the hh component of eq.~(\ref{eq:veltman_singlet-extension})]
smaller than 0.1, 0.5 and 1.0, respectively.}
\label{fig:veltman_region_plots}
\end{figure}

\subsection{Veltman's condition for the Higgs mass}

Veltman's condition is a condition for the quadratic
divergence
to vanish in the radiative correction to the Higgs potential.
Generally
the one-loop effective potential for the Higgs boson in a
cut-off regularization scheme is given by
\begin{align}
	V_\text{1-loop}(\phi)
=	\frac{1}{64\pi^2}\text{STr}
	\left[
		\varLambda^4
		\left(
			\ln\varLambda^2
			-\frac{1}{2}
		\right)
		+2M^2(\phi)\varLambda^2
		+M^4(\phi)
			\left(
				\ln\frac{M^2(\phi)}{\varLambda^2}
				-\frac{1}{2}
			\right)
	\right]
	+\text{c.t.}\,,
\end{align}
where $\varLambda$ is a UV regulator, and
$M^{2}(\phi) (\ll \varLambda^{2})$ is the mass-squared matrix .
The first term is a cosmological constant term, which we subtract by the counter term (with tremendous fine-tuning, but we will not be
concerned about it here).
The second term is a quadratically divergent term.
It can either be subtracted (fine-tuning), which makes the theory unnatural, or vanish if 
\begin{align}
	\frac{1}{2}\frac{\del^2}{\del (\delta \phi_{i})\del (\delta \phi_{j})}
	\left[
	\text{STr}~M^2
		\left(
			\braket{\vec{\phi}}
			+\delta\vec{\phi}
		\right)
	\right]_{\delta\vec{\phi}\rightarrow\vec{0}}
=	0
~~~\text{for each }i,\,j
\label{eq:veltman_def}
\end{align}
holds at a scale $\mu_0$, 
which makes the theory natural.
Here, $\delta \vec{\phi}$ denotes a fluctuation field vector around the VEV $\braket{\vec{\phi}}$\,.

In the SM, the coefficient of the quadratically divergent term is 
given by
\begin{align}
		\frac{1}{2}\frac{\del^2}{\del h^2}
	\left[
	\text{STr}~M^2
		\left(
			v_\text{H}+h
		\right)
	\right]_{\mu=v_\text{H},\,h\rightarrow0}
&=	
\frac{9 g^2(v_\text{H})}{4}+\frac{3 {g'}^2(v_\text{H})}{4}-6 y_\text{t}^2(v_\text{H})+6\lambda_\text{H}(v_\text{H})
\nonumber\\
&\simeq
	-4.1\,.
\label{eq:veltman_SM}
\end{align}
Thus, Veltman's condition is violated at order unity 
and fine-tuning is necessary if the cut-off scale
is much higher than the electroweak scale.

Here, we examine whether the fine-tuning is tamed comparatively
in the scale-invariant model.
The coefficient of $\varLambda^2$ is expressed by the matrix
\begin{align}
&	\frac{1}{2}\frac{\del^2}{\del (\delta\phi_{i})\del (\delta\phi_{j})}
	\left[
	\text{STr}~M^2
		\left(
			\braket{\vec{\phi}}+ \delta \vec{\phi}
		\right)
	\right]_{\delta \vec{\phi}\rightarrow\vec{0}}
\nonumber\\
&=
	\left(
	\begin{array}{l}
 		\frac{9 g^2(v_\text{H})}{4}
		+\frac{3 {g'}^2(v_\text{H})}{4}
		-6 y_\text{t}^2(v_\text{H})
		+6\lambda_\text{H}(v_\text{H})
		+N \lambda_\text{HS}(v_\text{H}) 
~~~~~~~~~~~~~~~~~~~~~ 0 
~~~~~~~~\\
~~~~~~~~~~~~~~~~~~~~~0 
~~~~~~~~~~~~~~~~~~~~~ ~~~~~~~~~~~~~~~
\rule{0mm}{6mm}
		(4 \lambda_\text{HS}(v_\text{H})+(N+2) \lambda_\text{S}(v_\text{H}) ){\bf 1}_{N\times N}
	\end{array}
	\right) .
\label{eq:veltman_singlet-extension}
\end{align}
It is diagonal, since there is no mixing between the
Higgs boson and the singlet scalar boson.
Let us moderately examine only the coefficient of the
Higgs boson for various $N$.
The hh component of the above matrix takes values 
{$0.76$, $5.6$, $12.8$}
for $N=1$, 4, 12, respectively.
Instead, we can examine the constraint by Veltman's condition on the 
parameters of the model, see Figs.~\ref{fig:veltman_region_plots}(a)-(c), 
which also combine the phenomenologically valid regions plotted
in  Figs.~\ref{fig:rpPheno}.
As can be seen, the fine-tuning is tamed particularly for $N=1$, which
may be a good feature of the model.\footnote{
Since the cut-off scale, given by the Landau pole,
is at several TeV scale in the $N=1$ case, there will not be
a serious
fine-tuning problem below this scale.
}

On the other hand,
the quadratic divergence for the singlet component is significant.
Although $\lambda_\text{S}$ is a free parameter, it is
expected to be positive semi-definite at the electroweak scale in order to
stabilize the vacuum at LO.\footnote{
Since one-loop effects induced by the
portal coupling are large, there may be a consistent region
where
$\lambda_\text{S}$ is negative and still the vacuum is stabilized.
}
One way to remedy this problem is to couple right-handed Majorana
neutrinos to the singlet scalars, as advocated in
\cite{Antipin:2013exa}.
Here, we do not pursue such possibilities and leave it as
an open question.

\section{Conclusions and Discussion}

Up to now experimental data show that properties of the Higgs boson are consistent with the SM predictions.
It is an intriguing question, with respect to the structure of the vacuum, whether only the Higgs self-interactions not measured so far can deviate significantly from the SM predictions.
A possible scenario is 
that the Higgs effective potential  is irregular at the origin.
In this analysis we studied an extension of the Higgs sector with classical scale invariance as an example of such potentials.
As a minimal model, we considered the SM Higgs sector in the scale-invariant limit, coupled via a portal interaction to an $N$-plet of real scalars under a global $O(N)$ symmetry (singlet under the SM gauge group).
We analyzed the electroweak symmetry breaking by the CW mechanism using RG and examined interactions among the scalar particles which probe the structure of the potential around the vacuum.

We computed the effective potential in the configuration space of the Higgs and singlet scalar fields $(\phi,\,\varphi)$.
The input parameters are the Higgs self-coupling $\lambda_\text{H}$, portal coupling $\lambda_\text{HS}$, and self-coupling of the singlet scalars $\lambda_\text{S}$ at tree level.
In order to obtain perturbatively valid predictions, we find that parametrically $|{\lambda_\text{H}}| \ll |{\lambda_\text{HS}}|$ needs to be satisfied.
Furthermore, $\lambda_\text{HS}$ needs to be large, of order $\sqrt{N_\text{C}/N}\,y_\text{t} \approx 1.7N^{-1/2}$, in order to overwhelm the  contribution of the top quark loop.
We developed a special perturbative formulation for our model in order to analyze the $(\phi,\,\varphi)$ space, since a naive treatment fails. 
The consistent parameter region of the couplings $(\lambda_\text{H},\,\lambda_\text{HS},\,\lambda_\text{S})$ is found which gives
a global minimum of the effective potential at $(\phi,\,\varphi)=(v_\text{H},\,0)$ with the Higgs VEV $v_\text{H}=246$~GeV and the Higgs boson mass $m_\text{h}=126$~GeV. 
The potential is stable against RG improvements, showing validity of the perturbative predictions.
$\lambda_\text{H}$ and $\lambda_\text{HS}$ are almost fixed, while $\lambda_\text{S}(>0)$ can be taken fairly freely.
The SM gauge group is broken as in the SM, while the $O(N)$ symmetry is unbroken. 
(Hence, there is no NG boson.)
The Higgs boson and singlet scalars do not mix at the vacuum.
The mass of the $N$ degenerate singlet scalars arises at LO of the perturbative expansion, whereas the mass of the Higgs boson is generated at NLO.
Hence, generally the singlet scalars are much heavier than the Higgs boson.
For $N=1$ and $\lambda_\text{S}=0.1$ the mass is about 500~GeV, and it becomes lighter for larger $N$ and heavier for larger $\lambda_\text{S}$.

Since the coupling $\lambda_\text{HS}$ should be large in order to beat the top loop contribution, the Landau pole appears at a few to a few tens TeV (the position of the Landau pole is higher for larger $N$ and smaller for larger $\lambda_\text{S}$). 
Hence, the cut-off scale of this model is considered to be around this scale.
For instance, this feature conflicts with a scenario which imposes a classically scale-invariant boundary condition at the Planck scale as a possible solution to the naturalness problem.
Even without such a motivation, it can be 
a serious drawback of this model that the Landau pole is located so close to the electroweak scale.
These features are consistent with the results of the analysis given
in \cite{Dermisek:2013pta}, while we presented more detailed analyses.

We computed the triple and quartic (self-)couplings of the Higgs and singlet scalar particles at the vacuum.
Computation of the interactions involving the
singlet scalars is a unique aspect of the use
of our formulation, since the portal coupling is large and
loop-induced effects tend to be large.
We obtain the Higgs triple and quartic self-couplings which are larger than the SM values by factors 1.6--1.8 and 2.8--4.5, respectively.
The triple coupling is hardly dependent on $N$, while the quartic coupling decreases slightly with $N$.
Both of them are barely dependent on $\lambda_\text{S}$.
According to the studies \cite{Baer:2013cma,Asner:2013psa}, we naively expect that the triple coupling can be detected at $2\sigma$ level at a future ILC, with an integrated luminosity of 1.1~ab$^{-1}$ at a centre-of-mass energy $\sqrt{s}=500$~GeV and 140~fb$^{-1}$ at $\sqrt{s}=1$~TeV.
The couplings involving the singlet scalars are fairly large, of order ten, in the case of small $N$, reflecting the irregularity of the potential at the origin and a large value of $\lambda_\text{HS}$.
If probed, these large couplings provide fairly vivid clues to the structure of the effective potential in the vicinity of the vacuum.

It is difficult to detect signals of this model at the current LHC experiments.
Since there is no mixing between the Higgs and singlet scalar particles, the couplings of the Higgs boson with other particles are unchanged from the SM values at tree level.
The singlet scalar particles interact with the SM particles only through the Higgs boson.
Furthermore, the singlet scalars are heavy and can be produced only in pairs due to the $O(N)$ symmetry (or $Z_2$ symmetry for $N=1$).
These features make it difficult to detect the singlet scalars directly at the LHC experiments.
The production cross sections of the singlet scalars at the LHC are expected to be very small and it would be difficult to detect them, although a further detailed analysis is needed.
On the other hand, it would be difficult to probe the extended Higgs sector from loop effects in various precision measurements.
These appear as higher-order effects to the Higgs effects, and since the Higgs effects themselves are small, we expect that detection of an anomaly is non-trivial.

From the cosmological point of view, there is a possibility that the singlet scalar boson(s) is a part of dark matter.
The singlet boson is stable due to 
$O(N)$ or $Z_2$ symmetry.
Since the couplings between singlet(s) and Higgs $\lambda_\text{hss}$ and $\lambda_\text{hhss}$ are very large, the annihilation cross section is also large, which decreases its relic abundance.
In $N>1$ cases, the total relic abundance is the sum over each individual singlet $\text{s}_i$.
Then typically the scalar couplings decrease and the total relic abundance increases with $N$.
Our naive estimation shows that singlet(s) can be a dark matter, 
whose abundance is less than around $1\%$ of the total dark matter relic abundance \cite{Cline:2013gha}.
We are currently preparing a further detailed study.

We also examined Veltman's condition (vanishing of
the quadratic divergence)
for the Higgs mass.
We find that for $N=1$ the fine-tuning is relaxed
compared to the SM,
by the effect of the large portal coupling which cancels against
the top-quark loop effects.
Since the current level of the fine-tuning in the SM indicates
that the natural scale of the cut-off is quite close to
the electroweak scale, this may be a good tendency of the
model.
The relaxation of fine-tuning is not significant for $N>1$.

A possible scenario to avoid the Landau pole near the electroweak scale is to promote the global $O(N)$ symmetry to a gauge
symmetry (the group can also be replaced by another non-abelian group).
An appropriate gauge-symmetric extension pushes up
the location of the Landau pole, driven by
the asymptotically-free nature, and in an extreme case, up to
the Planck scale in the context of a radiative electroweak symmetry breaking scenario \cite{Dermisek:2013pta}.

Since the Higgs self-interactions by higher powers of the Higgs field are not suppressed and scalar interactions are large, one may suspect that our model belongs to one of the strongly-interacting Higgs sector, which can be analyzed, e.g.\ using a non-linear sigma model. 
We note that it is not a strongly-interacting model, at least around the electroweak scale. 
We reemphasize that our predictions are well
within the perturbative regime,
and the usual loop expansion with only renormalizable interactions gives stable predictions with only a few
input parameters.
To realize a phenomenologically valid scenario, a slightly unusual order-counting is employed, 
as discussed in Sec.~3.

\section*{Acknowledgement}

The works of K.E. and Y.S. were supported in part by Tohoku University Institute for International Advanced Research and Education and by Grant-in-Aid for scientific research No. 23540281 from MEXT, Japan, respectively.

\bibliographystyle{utphys}
\bibliography{references}

\end{document}